\documentclass{article}
\pdfoutput=1
\PassOptionsToPackage{numbers, compress}{natbib}

\usepackage[preprint]{neurips_2021}

\usepackage[utf8]{inputenc} 
\usepackage[T1]{fontenc}    
\usepackage{hyperref}       
\usepackage{url}            
\usepackage{booktabs}       
\usepackage{amsfonts}       
\usepackage{nicefrac}       
\usepackage{microtype}      
\usepackage{xcolor}         
\usepackage{amsmath}
\usepackage[tight]{subfigure}
\usepackage{graphicx}
\usepackage{multirow}
\usepackage{amsthm}
\usepackage{caption}
\usepackage{wrapfig}
\usepackage{authblk} 

\title{PEEL: A Provable Removal Attack on \\ Deep Hiding}

\author[1]{Tao Xiang}
\author[1]{Hangcheng Liu}
\author[1]{Shangwei Guo}
\author[2]{Tianwei Zhang}
\affil[1]{College of Computer Science, Chongqing University, China}
\affil[2]{School of Computer Science and Engineering, Nanyang Technological University, Singapore}

\begin{document}

\maketitle

\begin{abstract}
Deep hiding, embedding images into another using deep neural networks, has shown its great power in increasing the message capacity and robustness. In this paper, we conduct an in-depth study of state-of-the-art deep hiding schemes and analyze their hidden vulnerabilities. Then, according to our observations and analysis, we propose a novel ProvablE rEmovaL attack (PEEL) using image inpainting to remove secret images from containers without any prior knowledge about the deep hiding scheme. We also propose a systemic methodology to improve the efficiency and image quality of PEEL by carefully designing a removal strategy and fully utilizing the visual information of containers. Extensive evaluations show our attacks can completely remove secret images and has negligible impact on the quality of containers.

\end{abstract}
\bibliographystyle{unsrt}

\section{Introduction} \label{sec:intro}
Data hiding \cite{SurveyDH} is the art of hiding secret data within a cover image or other multimedia signals imperceptibly
and has gained popularity in applications such as secret communication \cite{Stegastamp,VideoSte,LFM}, copy-right protection \cite{DMIP}, and content authentication \cite{Hidden}.
A data hiding scheme usually consists of two algorithms: \emph{hiding} and \emph{revealing} algorithms. The hiding algorithm encodes and embeds a secret message insider a cover image without affecting its visual perception that is outputted as the container. During the revealing phase, the revealing algorithm takes a container as input and reveals the secret message. Traditional data hiding schemes \cite{Hugo,Hill,Wow} can guarantee the revealed secret message is the same as the embedded one, but have a limited message capacity. For example, the capacity of HUGO \cite{Hugo}, a well-known data hiding scheme, is less than 0.5 bits per pixel (bpp).

\begin{figure}[t]
	\centering
	\subfigure[DDH]{\includegraphics[width=0.35\linewidth]{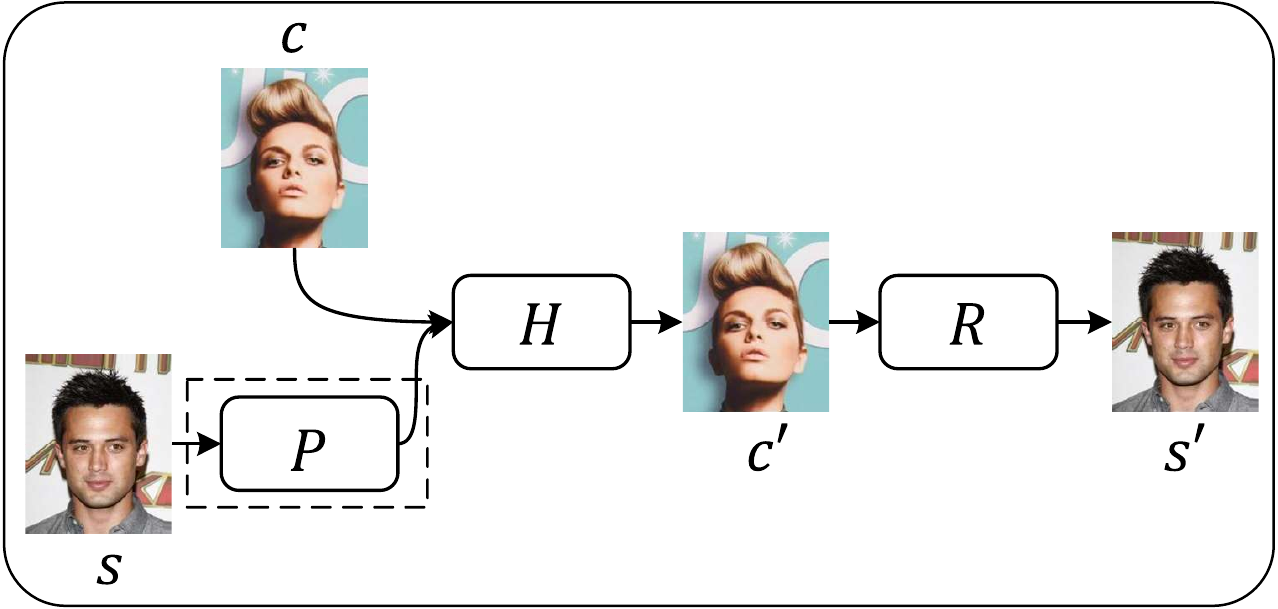}\label{fig:architectures_steganography:a}}
  \subfigure[UDH]{\includegraphics[width=0.595\linewidth]{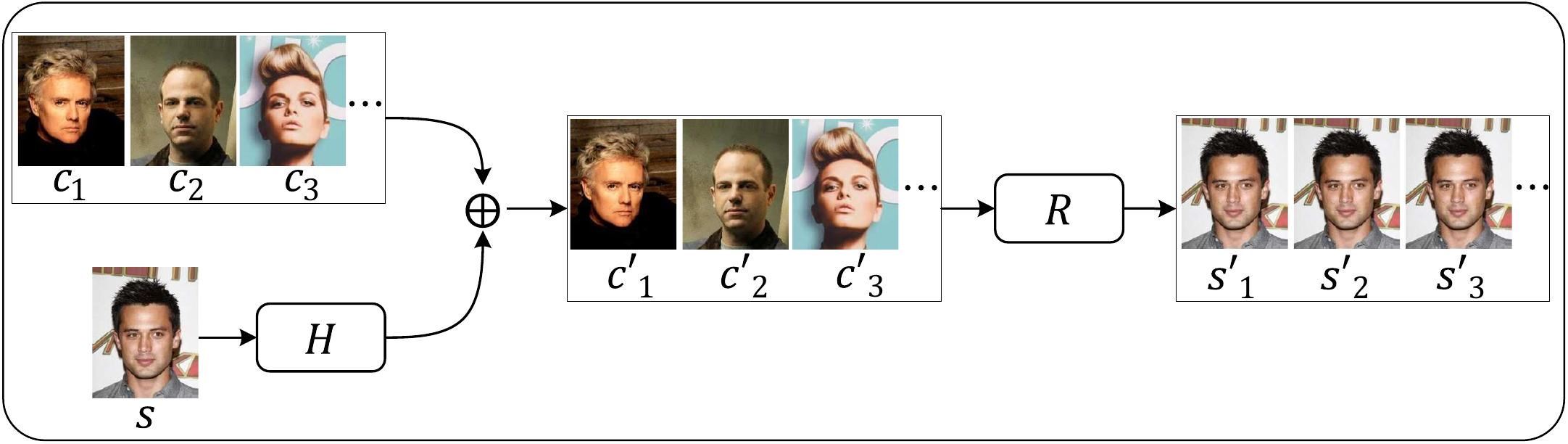}\label{fig:architectures_steganography:b}}
	\caption{Two kinds of meta-architectures for deep hiding, where $H, R$ are hiding and revealing models; $s$, $c$ ($c_i$) are secret and cover images; $c'$ ($c'_i$), $s'$ are container and revealed images; $P$ is an optimal preprocessing network; $\oplus$ is the element-wise addition.}
	\label{fig:architectures_steganography}
\end{figure}

Taking advantage of advance deep learning techniques, recent data hiding (i.e. deep hiding) schemes \cite{DS,ISGAN,UDH,HIWI} achieve a much higher message capacity (e.g. 24 bpp) and can hide a full image within another while preserving the visual quality of both secret and cover images. Existing deep hiding schemes can be classified into two categories based on their meta-architectures \cite{SurveyDH}: cover-dependent deep hiding (DDH) \cite{DS,ISGAN,HIWI} and universal deep hiding (UDH) \cite{UDH}. Figure \ref{fig:architectures_steganography} illustrates the meta-architectures of DDH and UDH. DDH encodes a secret image depending on the cover image, while UDH can encode and embed a secret image into multiple covers.

Although deep hiding schemes have greatly improved the message capacity, they have several fatal vulnerabilities in robustness. First, deep hiding schemes only embed pixels of secret images into the corresponding tiny regions of the cover image, which means the secret pixels cannot be revealed if the regions are destroyed. Second, the hiding is of low redundancy. Destroyed secret pixels cannot be recovered from or even affect the reveal of surrounding pixels.



Inspired by the above vulnerabilities, we propose a novel ProvablE rEmovaL attack (PEEL) on deep hiding to delete secret images from containers while preserving their visual quality. Our PEEL mainly consists of two phases: removal and inpainting. In particular,  at each iteration, we remove pixels in a small sub-region of a container and maintain the visual quality of the container using image inpainting techniques \cite{EdgeConn,GICIC,structFlow,LSIC}. We repeat this process until all pixels in the container are removed and repaired once. Compared with existing removal attacks \cite{Destruction, PixelSteganalysis} that either rely on prior knowledge about deep hiding schemes or fail to remove secret images, we theoretically prove that PEEL can remove secret images as well as preserve the visual quality of container images without any prior knowledge.

We also propose a systemic methodology to optimize PEEL, which consists two design strategies. First, for the removal phase, we deal with more sub-regions of a container at each iteration to improve the efficiency of PEEL. Second, for the inpainting phase, we provide extra information of the removed regions (e.g., edge and distorted pixels) to improve the quality of the repaired container. With the carefully designed strategies, the optimized PEEL can remove secret images within a constant number of iterations and highly improve the visual quality of containers. We conduct extensive experiments to show that our PEELs provide much stronger removal ability on state-of-the-art deep hiding schemes and their robustness-enhance versions.

\section{Background}
\subsection{Deep Hiding}
Deep hiding employs deep neural networks to improve the message capacity and robustness and consists of the hiding and revealing phases. A deep hiding scheme has two goals: (1) the hiding of secret images cannot affect the visual quality of covers; (2) the recovered image must be consistent with the embedded one. We can formally define deep hiding as below.
\newtheorem{definition}{Definition}
\begin{definition}($\xi$-deep hiding)
    Let $c$ and $s$ be cover and secret images with the same size. $H$ and $R$ are the hiding and revealing algorithms designed using deep neural networks, where $H(c, s)$ outputs a container image $c'$ and $s' = R(c')$. Let $D$ be a visual distance metric between two images. The scheme $\mathcal{S} = (H, R)$ is a $\xi$-deep hiding if for $\forall \ c$ and $s$, (1)  $\mathbb{E}_{\sim s,c}D(c,c') \leq \xi$ and (2) $\mathbb{E}_{\sim s,c}D(s,s') \leq \xi$.
    \label{def:hiding}
\end{definition}

With deep hiding, a secret image can be encoded and embedded into a cover image imperceptibly. Since secret images are the same size with covers and carry a huge amount of information, the encoding and embedding strategies are critical for a deep hiding scheme. As shown in Section \ref{sec:intro}, existing deep hiding schemes can be classified into two categories, DDH and UDH, according to the difference between the hiding strategies.

\textbf{DDH.} For a secret image, both the encoding and embedding processes dependent on a specific cover image. For example, Baluja et al. proposed a scheme DS \cite{DS}, in which they utilize a preprocessing model $P$ to encode secret images first. Then, given a cover image $c$, the hiding model encodes and embeds the output of $P$ into $c$. Zhang et al. proposed ISGAN \cite{ISGAN}, in which the hiding model directly takes $s$ and $c$ as its inputs for encoding and embedding. In general, the hiding phase of DDH can be described as $c'=H(c,P(s))$ or $c'=H(c,s)$, shown as Figure \ref{fig:architectures_steganography:a}

\textbf{UDH.} The encoding and embedding processes are independent to cover images in UDH. For example, Zhang et al. proposed a UDH scheme that assigns the same secret image to multiple covers and minimizes the difference between the revealed images of the corresponding containers during the training process. Thus, the hiding model can learn universal code maps for secret images. In the embedding process, the universal code maps can be embedded into multiple covers by element-wise addition, as shown in Figure \ref{fig:architectures_steganography:b}.

\textbf{Adversarial Training.} These deep hiding models can be further enhanced by adversarial training \cite{UDH,DMIP,Hidden} to improve their robustness to different image distortions. With adversarial training, hiding models add one or multiple distortion layers to distort containers after embedded secret images. Then revealing models are trained to reveal the secret images from both the original and distorted containers.
For example, Zhang et al. \cite{UDH} adversarially train deep hiding networks by adding one or multiple types of distortions (e.g. noise, blur, compression, etc) into containers and the adversarially trained deep hiding schemes can successfully reveal secret images from distorted containers.

\subsection{Vulnerabilities of Deep Hiding}
Although current deep hiding schemes have different meta-architectures, they have similar vulnerabilities on robustness, which are summarized as follows. Although the robustness of existing data hiding schemes can be enhanced by \textit{adversarial training} \cite{UDH,DMIP,Hidden}, the vulnerabilities still exist even with such techniques (see Supplementary).

\begin{wrapfigure}{r}{0.59\linewidth}
    \centering
      \begin{tabular}{c}
        \includegraphics[width=0.51\linewidth]{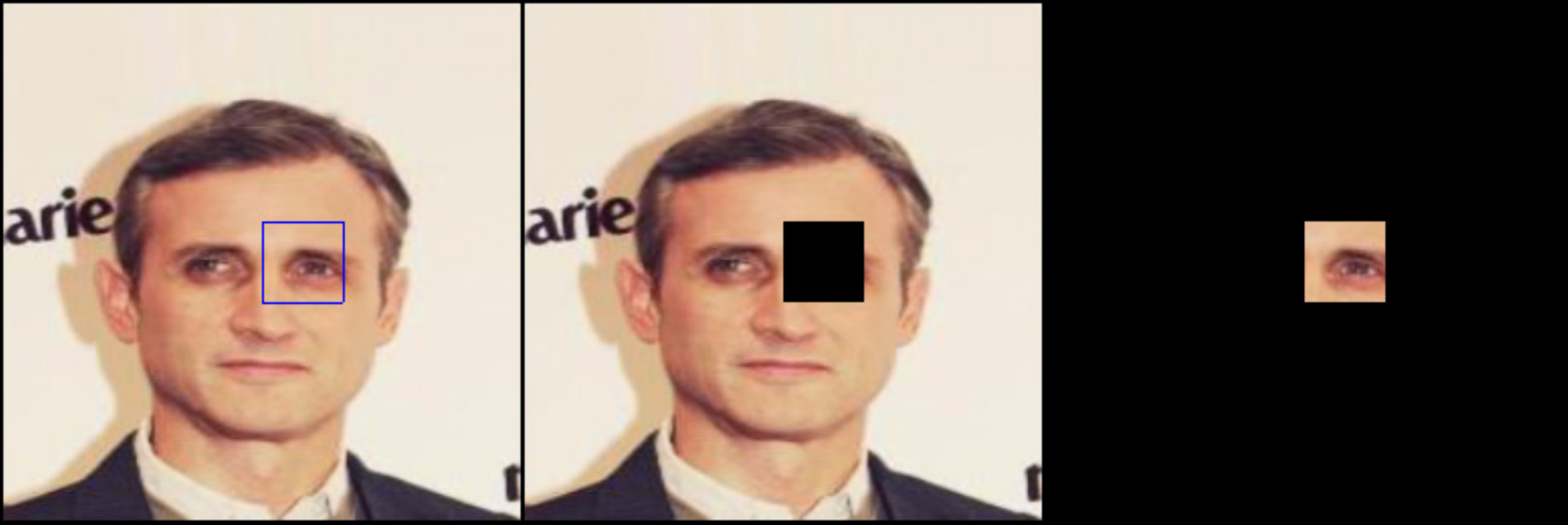}\\
        \includegraphics[width=0.51\linewidth]{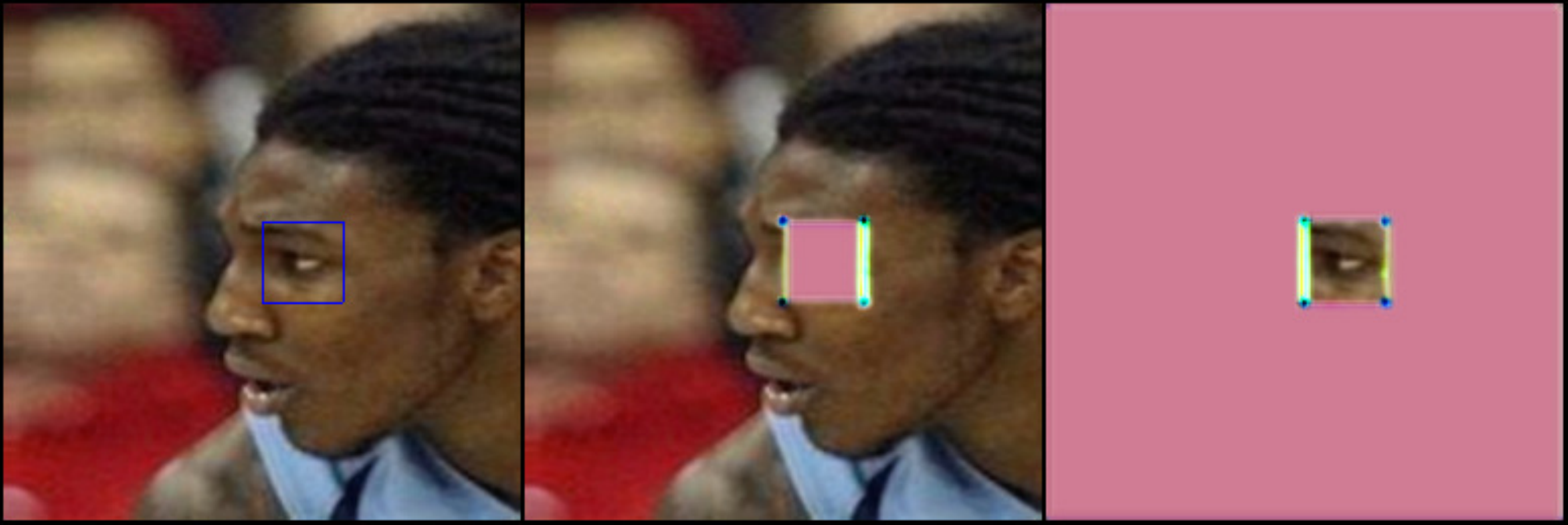} \\
        (a) Locality
      \end{tabular}
    %
      \begin{tabular}{c}
        \includegraphics[width=0.34\linewidth]{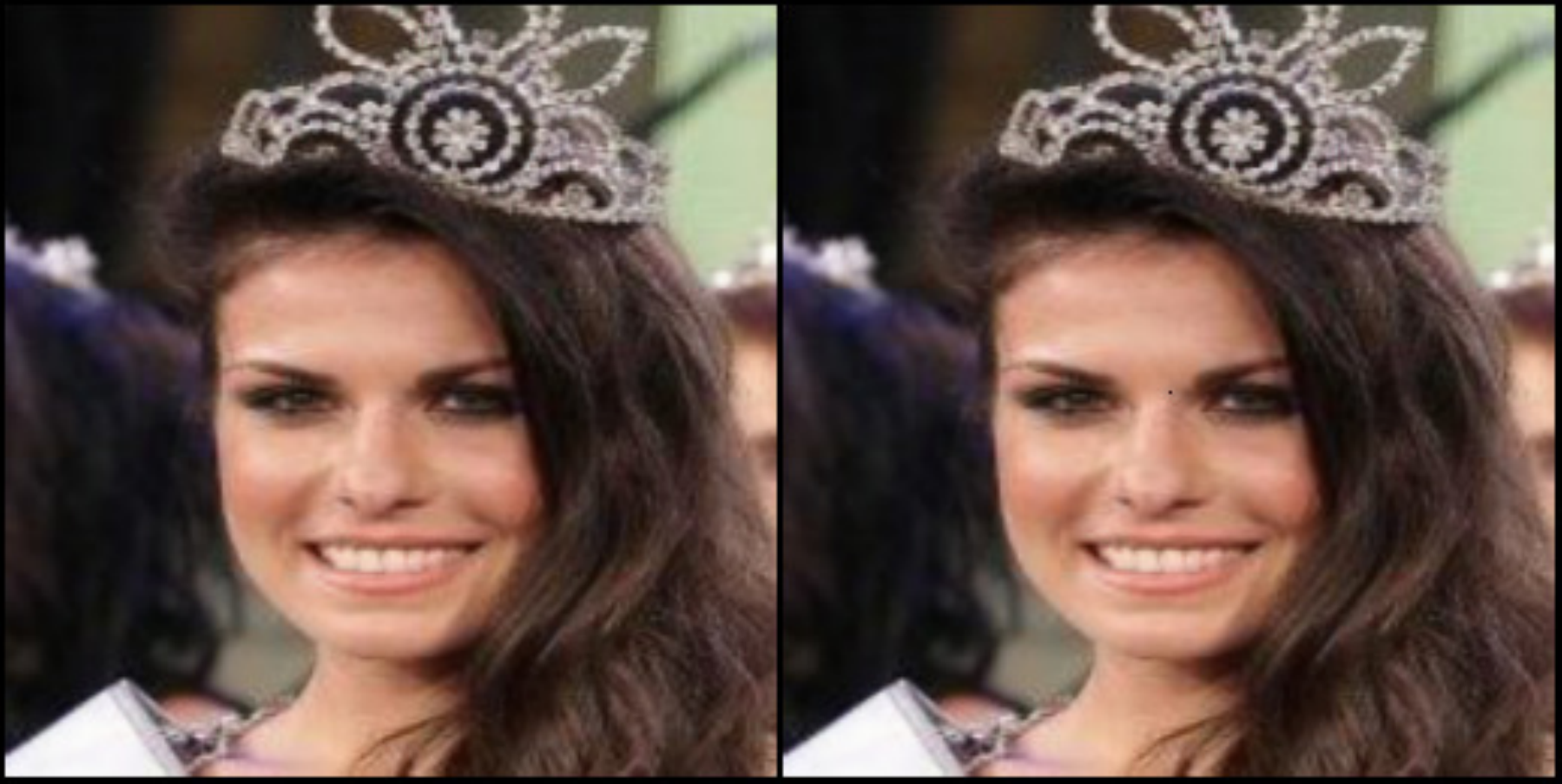}\\
        \includegraphics[width=0.34\linewidth]{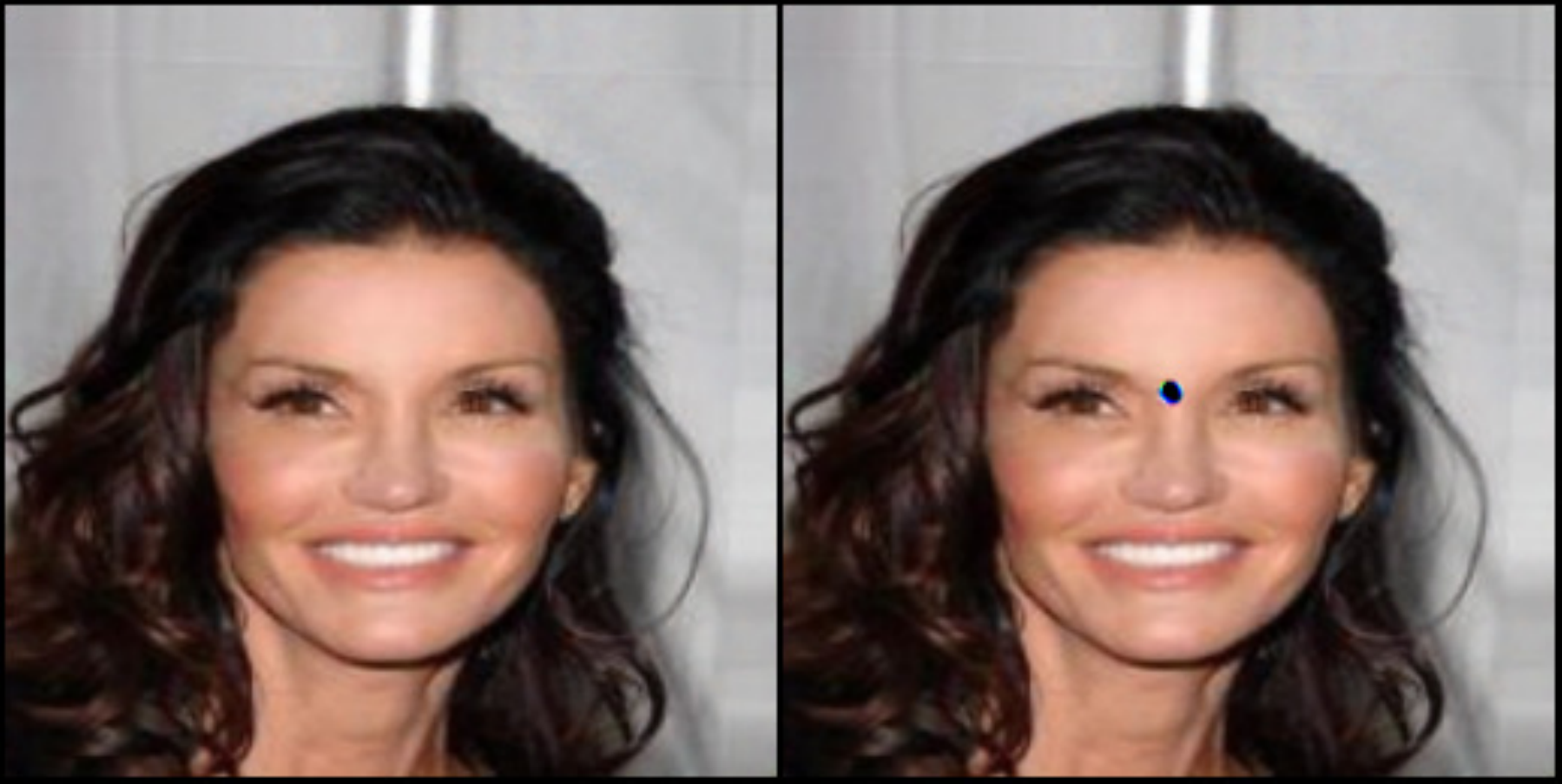} \\
        (b) Low redundancy
      \end{tabular}
    \caption{Vulnerabilities of deep hiding. The top images of each group are containers. The bottom are the corresponding revealed images. (a) We remove a random square region from the container (middle) and only keep the random region as the container (right); (b) We set one pixel of the right container image as 0.}
    \label{fig:vulnerabilities}
\end{wrapfigure}

\textbf{Locality.} For each pixel $s_{ij}$ of a secret image $s$, existing deep hiding embeds $s_{ij}$ in a tiny region of the corresponding pixel $c'_{ij}$ of the container $c'$. Figure \ref{fig:vulnerabilities} (a) illustrates the locality vulnerability of a well-known deep hiding scheme DS \cite{DS}. In Figure \ref{fig:vulnerabilities} (a), the last two pairs of the container and recover images show that if we remove one region of the container, the corresponding region of the recover image cannot also be revealed. Besides DS, other deep hiding schemes also have the vulnerabilities (see Supplementary). Thus, once some regions of containers are destroyed, it is impossible to reveal the corresponding secret pixels.

\textbf{Low Redundancy.}
This vulnerability means that if we only remove one pixel $c'_{ij}$, existing deep hiding schemes cannot recover a tiny region of the corresponding secret image. This phenomenon can be empirically proved in Figure \ref{fig:vulnerabilities} (b). In Figure \ref{fig:vulnerabilities} (b), we set one pixel of the top right container as 0, which causes a corresponding tiny region cannot be recovered. We also observe that UDH \cite{UDH} seems to be robust when setting the pixel as 0 but still has this vulnerability if we change the pixel to other pixel values (see Supplementary). This phenomenon also confirms that a secret pixel is embedded into a tiny region of the corresponding pixel in the container.

Due to the above vulnerabilities, it is possible to remove the hidden images from the container without affecting its visual perception. We define such removal attack as the following.
\begin{definition}(($\epsilon, \lambda$)-Removal Attack) \label{def:attack}
  Let $F$ be a removal attack on deep hiding. $\mathcal{S} = (H, R)$ is a deep hiding scheme. $F$ is a $(\epsilon, \lambda)$-removal attack on $\mathcal{S}$ if $\forall \ c$ and $s$, $\mathbb{E}_{\sim s,c}D(c', F(c'))\leq \epsilon$ and $\mathbb{E}_{\sim s,c}D(s', R(F(c'))\geq \lambda$.
\end{definition}
From Definition \ref{def:attack}, a removal attack has two goals: (1) the removal of secret images cannot affect the visual quality of containers; (2) the visual distance should be large enough between the revealed images before and after the removal attack. We will design a novel attack to achieve the two goals.

\section{Methodology} \label{sec:method}
Due to the above vulnerabilities, the robustness of deep hiding is challenged by this paper. We propose a ProvablE rEmovaL attack (PEEL) on deep hiding using image inpainting to remove secret images as well as preserve the visual quality of containers.
\subsection{PEEL}
\textbf{Insight.} Our PEEL mainly consist of two phases: removal and inpainting phases, which are used to remove secret images and preserve the quality of containers, respectively. Our design strateges are two-folds. \emph{First}, according to the locality vulnerability of deep hiding, secret pixels are encoded and embedded into the small regions of the corresponding pixels of a cover. Thus, to remove the secret pixels, we only need to remove the pixels in the corresponding regions of the container in the removal phase. However, the quality of the container drops sharply as we directly remove the pixels. Our \emph{second} design strategy is to utilize image inpainting techniques for recovering the missing regions. Since existing deep hiding schemes have the low redundancy vulnerability, the inpainting process would not contain the secret information of the missing regions.

\begin{figure}[t ]
	\centering
	\subfigure{\includegraphics[width=0.7\linewidth]{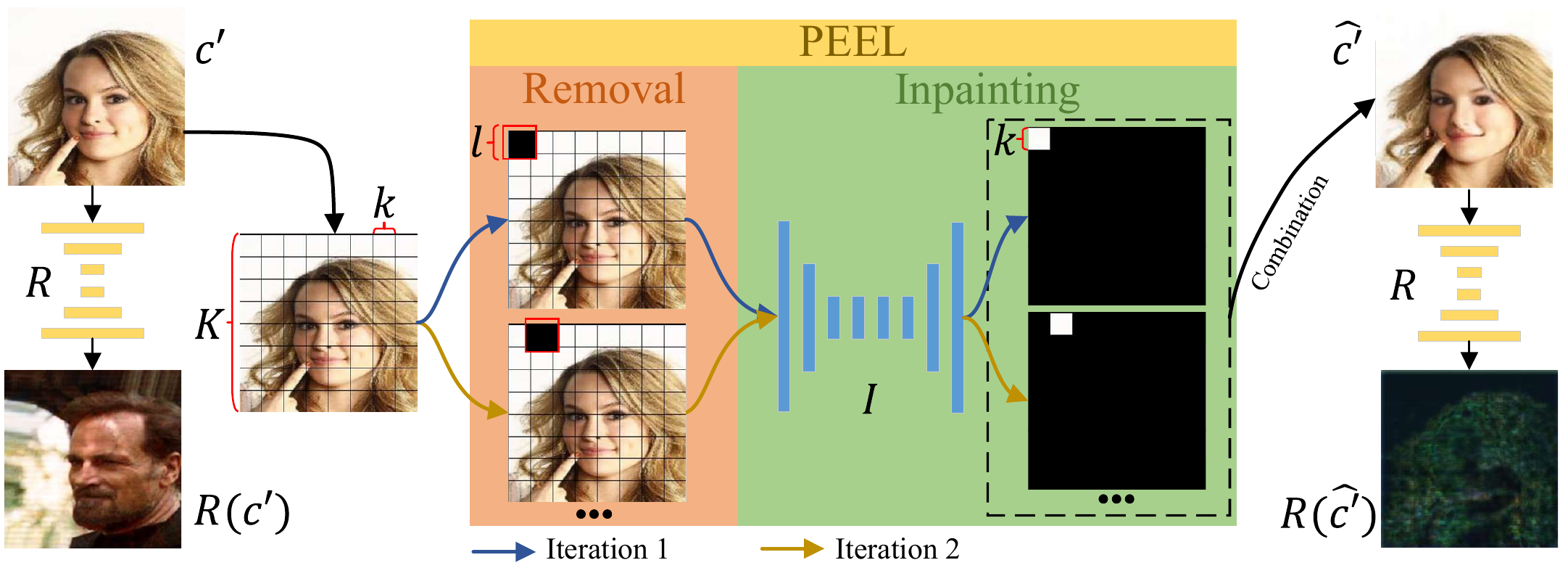}}
	\caption{Pipeline of PEEL. We divide $c'$ into square regions of side length $k$. We process one square region in each iteration of PEEL. Specifically, in the removal phase, we use a red box of size $l\times l$ ($l>k$) to determine the missing region, whose center coincides are the same with that of the chosen square region. We set all pixels in the red box as 0 to remove the content. In the inpainting phase, we complete the missing content with the pre-trained inpainting model $I$ and save the repair content. After processing all square regions, we combine all repair content as $\hat{c'}$.}

  \label{fig:peel}
\end{figure}

\textbf{Attacking Sketch.} We first train an image inpainting model $I$ using publicly available unlabeled images. Then, Given a container $c'$ of size $K\times K$, we first split $c'$ into small square regions with side length $k$, as shown in Figure \ref{fig:peel}. Then we remove and repair each square region. \emph{In the removal phase},
we set pixels of the $l\times l$ ($l > k$) region around $c'_{i,j}$ as 0. Note that we not only remove the square regions of side length $k$ but also remove its surrounding pixels, which ensures the secret information of the corresponding $k\times k$ region is removed completely.
The removal process can be described as
\begin{equation}
  c'^M=Removal(c',m,n,k, l),
\end{equation}
where $(m,n)$ are the coordinates of the center of current square region. $c'^M$ is the output of the removal function, in which the $l\times l$ square region with the center ($m, n$) is removed. \emph{In the inpainting phase}, we repair the missing region with $I$. It is worth noting that the hyper-parameters $k, l$ should be carefully chosen because a large $l$ would affect the quality of the repaired container. We finish our attack by repeating the above phases till all squares are removed and repaired once.

\textbf{Theoretical Analysis.} We theoretically analyze that for an arbitrary deep hiding scheme, our PEEL can completely remove secret images from containers under some reasonable assumptions. We first formalize the image inpainting technique.
\begin{definition}
  (\textit{($l$, $\gamma$)-Image Inpainting}) Let $M$ be a $K*K$ binary mask, in which elements in a square with side length $l$ are 0 and others are 1.
  Let $x$ is an image with the size of $K*K$. $I$ is a ($l$,$\gamma$)-image inpainting scheme if $\forall \ x$ and $M$, $\mathbb{E}_{\sim s,c}D(x,I(x\otimes M))\leq \gamma$, where $\otimes$ is the element-wise multiplication.
\end{definition}

We assume the visual metric is a norm and the locality and low redundancy vulnerabilities of deep hiding exist.
\newtheorem{assumption}{Assumption}
\begin{assumption}\label{ass:metric}
  The visual metric $D$ is a norm on the $K\times K$ image space.
\end{assumption}
\begin{assumption}\label{ass:hiding}
  Let $\mathcal{S} = (H, R)$ be an arbitrary deep hiding scheme. For $\forall \ s$ and $c$, $H$ hides each pixel $s_{ij}$ of $s$ independently within the corresponding square with the center $c_{ij}$ and side length $\frac{l-k}{2}$.
\end{assumption}

With Assumption \ref{ass:metric}-\ref{ass:hiding}, we can prove that our PEEL is a $(\epsilon, \delta)$-removal attack. The proof of Theorem \ref{theorem} can be found in Supplementary.
\newtheorem{theorem}{Theorme}
\begin{theorem}\label{theorem}
  Let $\mathcal{S} = (H, R)$ be an arbitrary $\xi$-deep hiding. The proposed PEEL is a $(\epsilon, \lambda)$-removal attack on $\mathcal{S}$ if $\gamma \leq \frac{\epsilon}{(\frac{K}{k})^2-1}$.
\end{theorem}

\subsection{Optimization Strategies}
Although PEEL can completely remove embedded secret images, there are two drawbacks: \textit{ low efficiency of inpainting} and \textit{low quality of containers}. First, removing and repairing regions one by one is inefficient. We have to iterate $(K/k)^2$ times to finish the attack. As the resolution increases, the efficiency problem becomes more serious. Second, due to the limitation of existing image inpainting techniques, the quality of repaired regions is low when the missing square regions cannot provide any visual information. To solve the above two drawbacks, we propose a systemic methodology to optimize and enhance the efficiency and quality of our PEEL, which includes the following two design strategies.

\textbf{Strategy 1: Efficiency Optimization.}
We deal with multiple regions at each iteration instead of only one. Roughly speaking, the surrounding pixels of a missing region highly influence the inpainting process. The influence decays as the distance to the missing region increases.
Inspired by this fact, we choose and remove multiple regions of containers at each iteration and the number of regions is proportional to the resolution of the containers. For example, as shown in Figure \ref{fig:opti_peel}, we select and remove a region in every four regions. Thus, we can fix the number of iterations to a constant, regardless of the resolution of containers.

\textbf{Strategy 2: Quality Optimization.}
We optimize the inpainting phase to improve the quality of cover images from two aspects. First, similar to \cite{EdgeConn} that predicts the edge of missing regions before the inpainting, we directly provide the real edge to the inpainting model. This edge information contains little secret information and can delineate and determine the contour of the missing regions. Second, we provide the inpainting model heavily distorted version of the missing regions, in which secret images are destroyed. Such optimization can provide more useful information for image inpainting and improve the quality of the repaired regions.

\subsection{Overall Attacking Process}
\begin{figure}[t]
	\centering
	\subfigure{\includegraphics[width=0.65\linewidth]{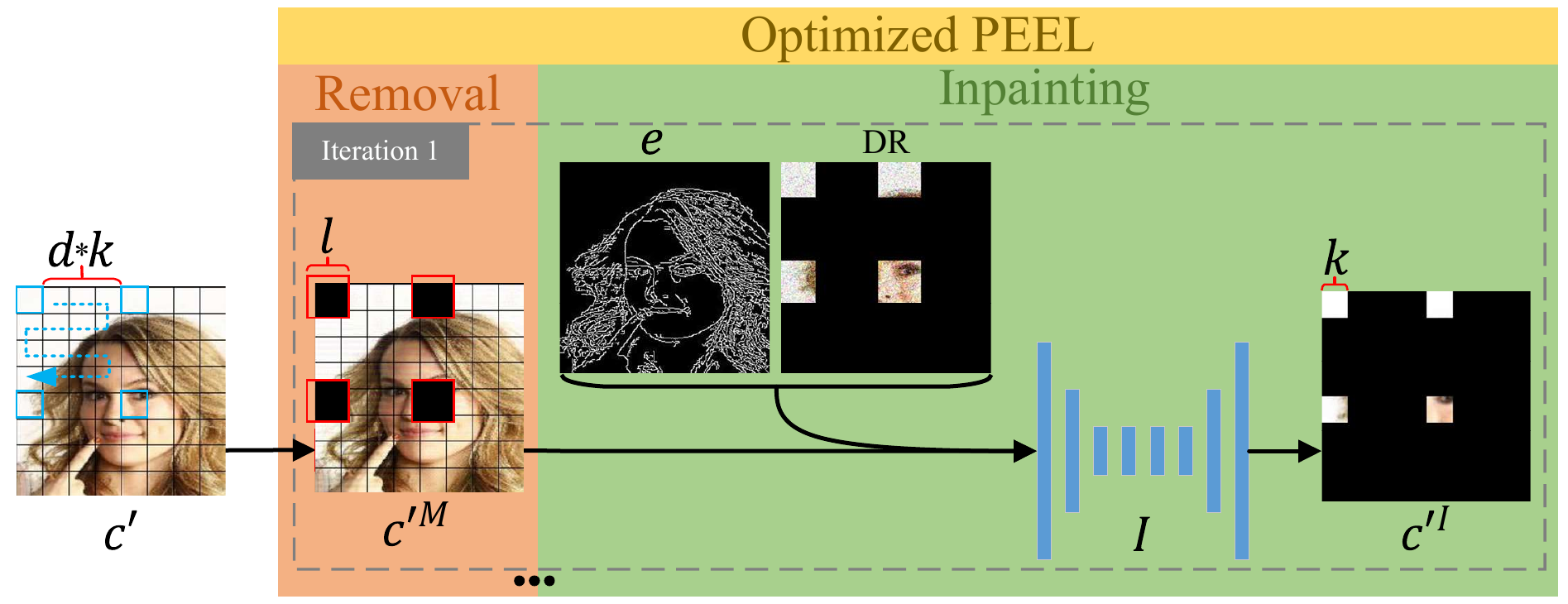}}
	\caption{Pipeline of the optimized PEEL. We process more square regions of size ($k\times k$), liking the regions marked by blue boxes, in each iteration. The removal phase is similar to that of PEEL except we use more red box to determine the missing regions. In the inpainting phase, we provide extra inputs to the inpainting model: the edge information of $c'$ and the seriously distorted version of the original pixels of the missing regions (DR). The edge is extracted by existing edge detectors and DR is produced through adding noise or other distortions. We save the repair content of all chosen square regions. Afterwards, we slide the blue boxes to choose new square regions. the of the blue boxes (e.g. the top left one) only need to slide following the blue path (the dotted line). After processing all square regions of $c'$, we combine all the repair content as $\hat{c'}$ of .}
    \label{fig:opti_peel}
\end{figure}

Taking advantage of the two optimization strategies, we describe the overall attacking process. We first prepare a well-trained image inpainting model.
In the removal phase, we deal with multiple sub-regions each time to remove the corresponding secret information. In the inpainting phase, we repair the missing region in high quality using the quality optimization strategy. The overall pipeline of the optimized PEEL is illustrated in Figure \ref{fig:opti_peel}.

\textbf{Removal Phase.}
Similar to PEEL, we divide $c'$ into square regions with the side length $k$ before the removal phase. The difference is that we choose multiple regions each iteration according to the efficiency optimization. We select a region every $d$ regions as marked by the blue boxes in Figure \ref{fig:opti_peel}. In the removal phase, we set pixels of all chosen regions and pixels around all chosen regions as 0. The pixels to be repaired are determined by the red boxes of size $l\times l$ ($l>k$). We denote the output of the removal phase as $c'^M$.

\textbf{Inpainting Phase}.
With the quality optimization, we provide the inpainting model the edge information ($e$) and the heavily distorted version of all missing regions (DR). $e$ can be extracted using existing edge detectors and the missing regions can be obtained by adding heavy distortions. Note that $e$ does not provide any information about the secret image. As for the DR, since we add heavy distortions, the secret information of the missing regions is destroyed and the revealing model can hardly recover secret pixels from the repaired regions. We complete the mission regions using the inpainting model, i.e.,
\begin{equation}
  c'^I=I(c'^M,e,\text{DR}).
\end{equation}

In the next iteration, we slide all blue boxes on $c'$ in the same direction to choose new regions. With the blue path in Figure \ref{fig:opti_peel}, we only conduct the removal and inpainting phases $(d+1)^2$ times to cover all regions of $c'$, no matter what the solution of $c'$ is. We collect the pixels of the repaired regions with size $k\times k$ in each iteration and own a new image without the secret image $\hat{c'}$.

\textbf{Preparation of the Inpainting Model.} We need to prepare a well-trained image inpainting model before the attacking process. The training dataset can be easily obtained by collecting cheap, publicly available, unlabeled images. We employ the loss function in \cite{EdgeConn} to train the inpainting model, i.e.,
\begin{equation}
  \min \mathcal{L}(\mathcal{L}_{recon}, \mathcal{L}_{perce}, \mathcal{L}_{style}, \mathcal{L}_{adver}),
  \label{eq:loss}
\end{equation}
where $\mathcal{L}_{recon}$, $\mathcal{L}_{perce}$, $\mathcal{L}_{style}$, and $\mathcal{L}_{adver}$ are reconstruction loss, perceptual loss, style loss, and adversarial loss, respectively. Details of these losses can be found in \cite{EdgeConn,StyleTransfer}.

For each training step, we first choose a batch of images from the training dataset. For each image, we divide it into square regions with the side length $k$. Then, we randomly select and remove regions as in the removal phase and provide the edge and distorted version of the removed regions for the minimization of $\mathcal{L}$.
Since these training images do not contain any secret images, the well-trained $I$ would not reveal secret information from the mission containers during the inpainting phase.

\section{Experimental Evaluation}
\subsection{Experiment Setup} \label{sec:exp:setup}
\textbf{Dataset.} Our attack is model-agnostic and dataset-agnostic. We conduct all experiments on the CelebA dataset \cite{CelebA}. We randomly select 80000 images to train the image inpainting models and 2000 images from the remain images for testing. One thousand of the 2000 images are used as cover images and the other thousand are used as secret images. All images are resized to the resolution of 255$\times$255 and the pixel values are normalized to the range of $[0,1]$.

\textbf{Deep Hiding.} We use state-of-the-art deep hiding schemes for evaluation, including DS \cite{DS}, ISGAN \cite{ISGAN}, and UDH \cite{UDH}.
Different from DS and UDH that embed color images into another, ISGAN hides gray images into color images because it only hides secret images in the Y channel of the YCrCb color space.

Further, we also consider the usage of adversarial training techniques to increase the robustness of the three schemes. In particular, we use a pre-trained auto-encoder to enhance the robustness.
The robust-enhanced deep hiding schemes are denoted as DS-AE, ISGAN-AE, and UDH-AE, respectively. We show the architecture of the auto-encoder in Supplementary.

\textbf{Implementation of PEELs and Baselines.}
For PEEL, we set $k=25$ and $l=35$.
Since the resolution 256 is not divisible by 25, we increase the size of containers by padding 0.
Before the inpainting, we remove all padded pixels to restore the original size.
For the optimized PEEL (PEEL-O), we set $k=50$, $d=2$, and $l=60$. Besides, we produce DR by adding scaled Gaussian noise to the original content of all missing regions. The Gaussian noise is sampled from the standard orthonormal distribution and is scaled by a real number $\delta$. The inpainting models $I$ in PEEL and PEEL-O have the same architecture (see Supplementary).

In addition to our proposed PEEL and PEEL-O, we also implement three other removal attacks for comparison:
\begin{itemize}
  \item \textbf{Gaussian Noise (GN)}. This removal attack adds Gaussian noise to the whole containers. Adding enough noise can prevent the recover of valid information of secret images. But, adding too much noise seriously reduces the usability of containers. We scale the noise using the $\delta$ to control the usability.
  \item \textbf{Gaussian Blurring (GB)}. One can use Gaussian blurring to remove secret images. In experiments, we set the kernel size and standard deviation as 5 and 3, respectively.
  \item \textbf{Median Blurring (MB)}. Similar to GB, we use the median blurring as a removal attack. And we set the kernel size as 5.
\end{itemize}
The architectures of all models involved in our experiments (e.g. the deep hiding models, the inpainting model, and the auto-encoder) and related configurations in training are detailed in Supplementary. We do not consider the two removal attacks \cite{PixelSteganalysis,Destruction} in our experiments for the follow two reasons. First, training the purifier in \cite{Destruction} needs much prior knowledge of the data hiding schemes. Second, as shown in the supplementary \footnote{https://anonymous-steganalysis.github.io} of \cite{PixelSteganalysis}, the proposed removal attack cannot remove secret images even without adversarial training.


\textbf{Metric.} To evaluate the performance of removal attacks, we measure the visual distance between two images using
PSNR and VIF \cite{VIF}.
A larger score of PSNR or VIF indicates a better content similarity between two images. PSNR is a wildly used metric due to its simplicity and mathematical convenience, but it is too simple to build a strong correlation between PSNR and the perception of the human visual system in most scenarios \cite{VIF,PVS}. Thus, we further choose VIF because it is design to quantify the information that is shared between the test and the reference images, which fits the scenario in this paper. In order to distinguish the two kinds of distances easily, we use PSNR-C and VIF-C to denote the distance between the cover images before and after attacks. And we use PSNR-S and VIF-S denote the distance between the revealed images before and after attacks.

\subsection{Overall Evaluation}
\begin{figure}[t]
  \centering
  \resizebox*{\textwidth}{!}{
      \begin{tabular}{cccccccc}
        & &GN(0.03)  &  GN(0.05) &GB&MB&  PEEL & PEEL-O \\
        \rotatebox{90}{$\ \ \ \ \ \ \ \ \ \ \ \ \ \  $UDH}&  \includegraphics[width=0.12\linewidth]{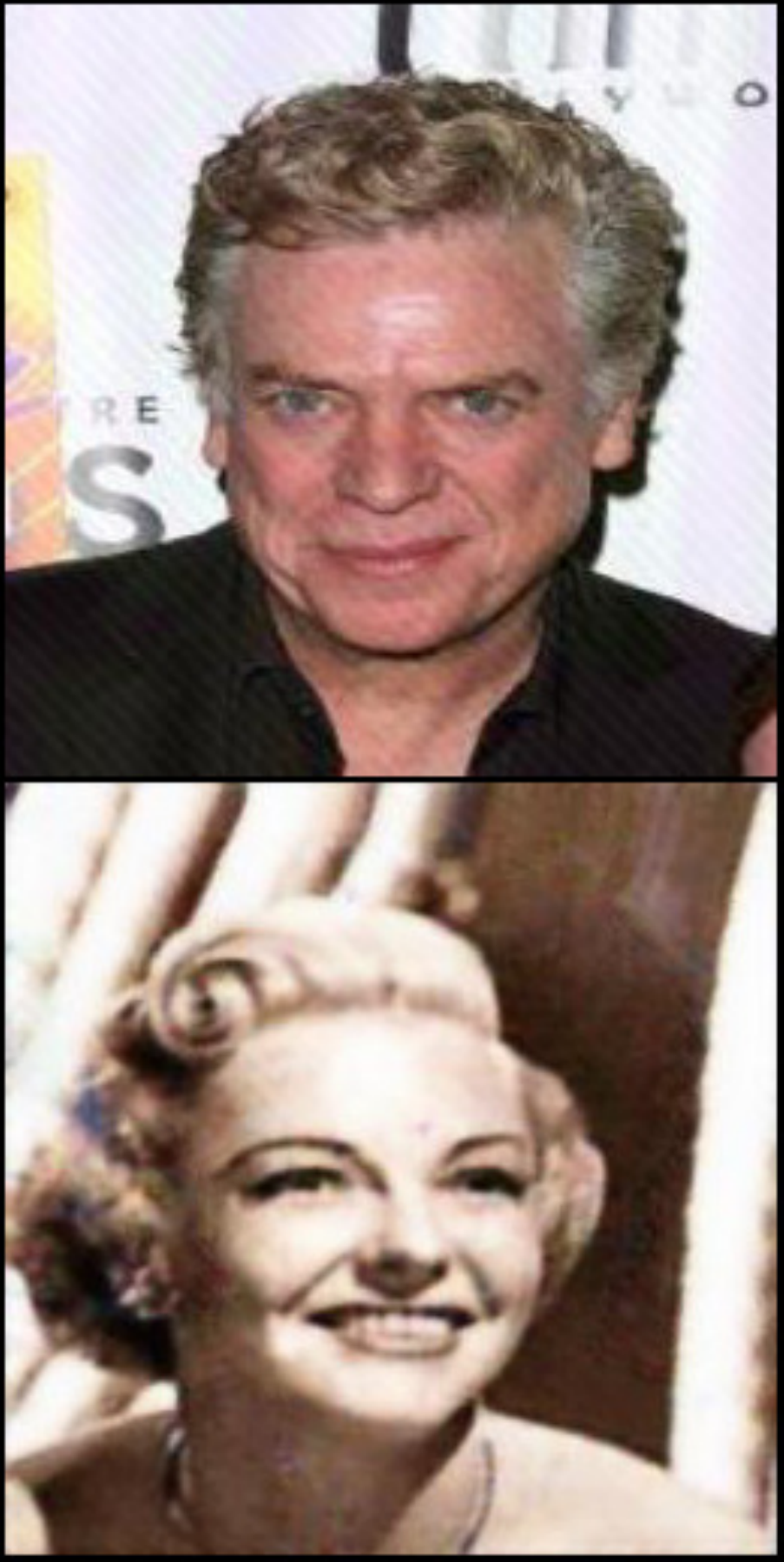}
        &\includegraphics[width=0.12\linewidth]{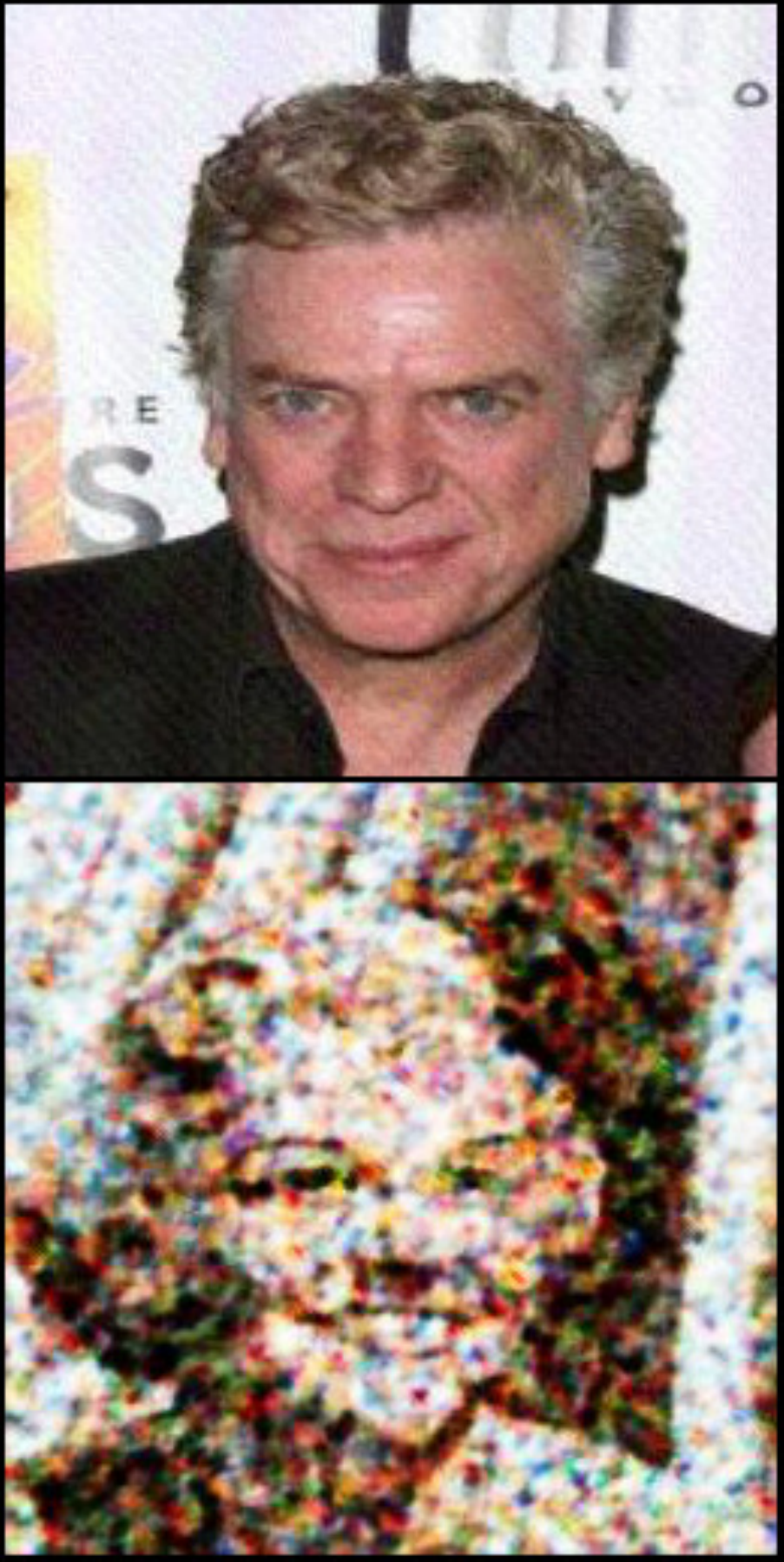}
        &\includegraphics[width=0.12\linewidth]{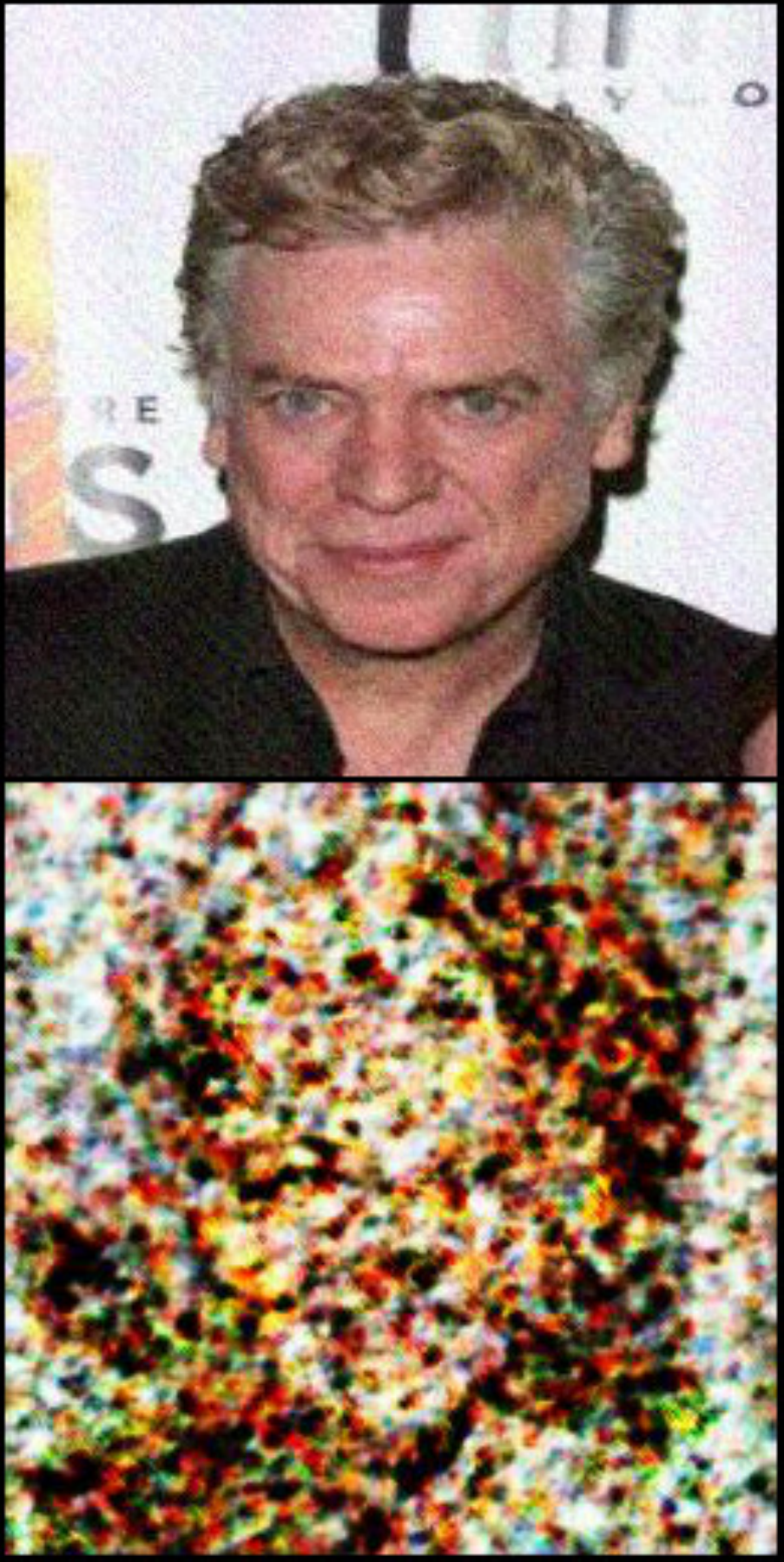}
        &\includegraphics[width=0.12\linewidth]{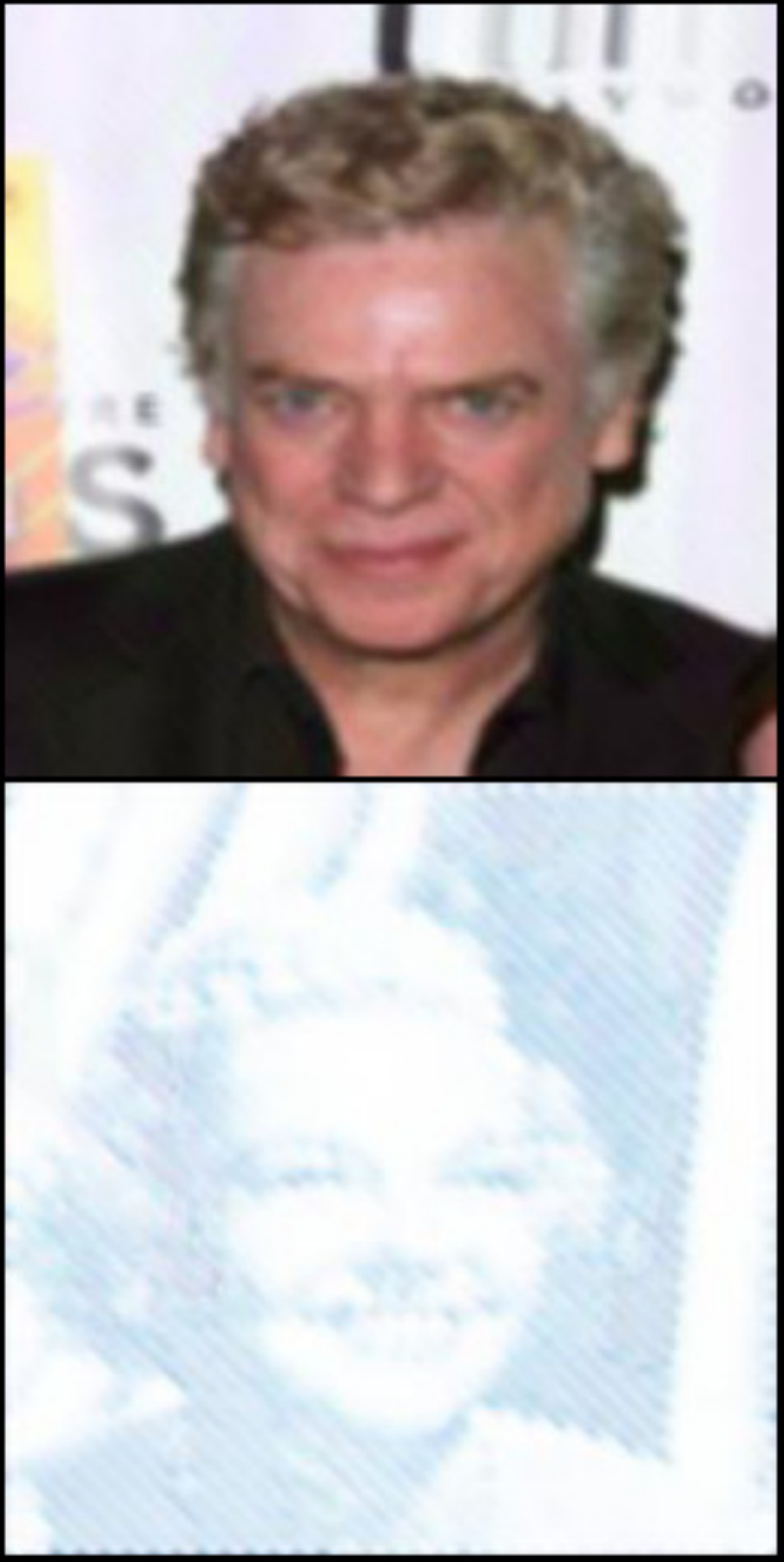}
        &\includegraphics[width=0.12\linewidth]{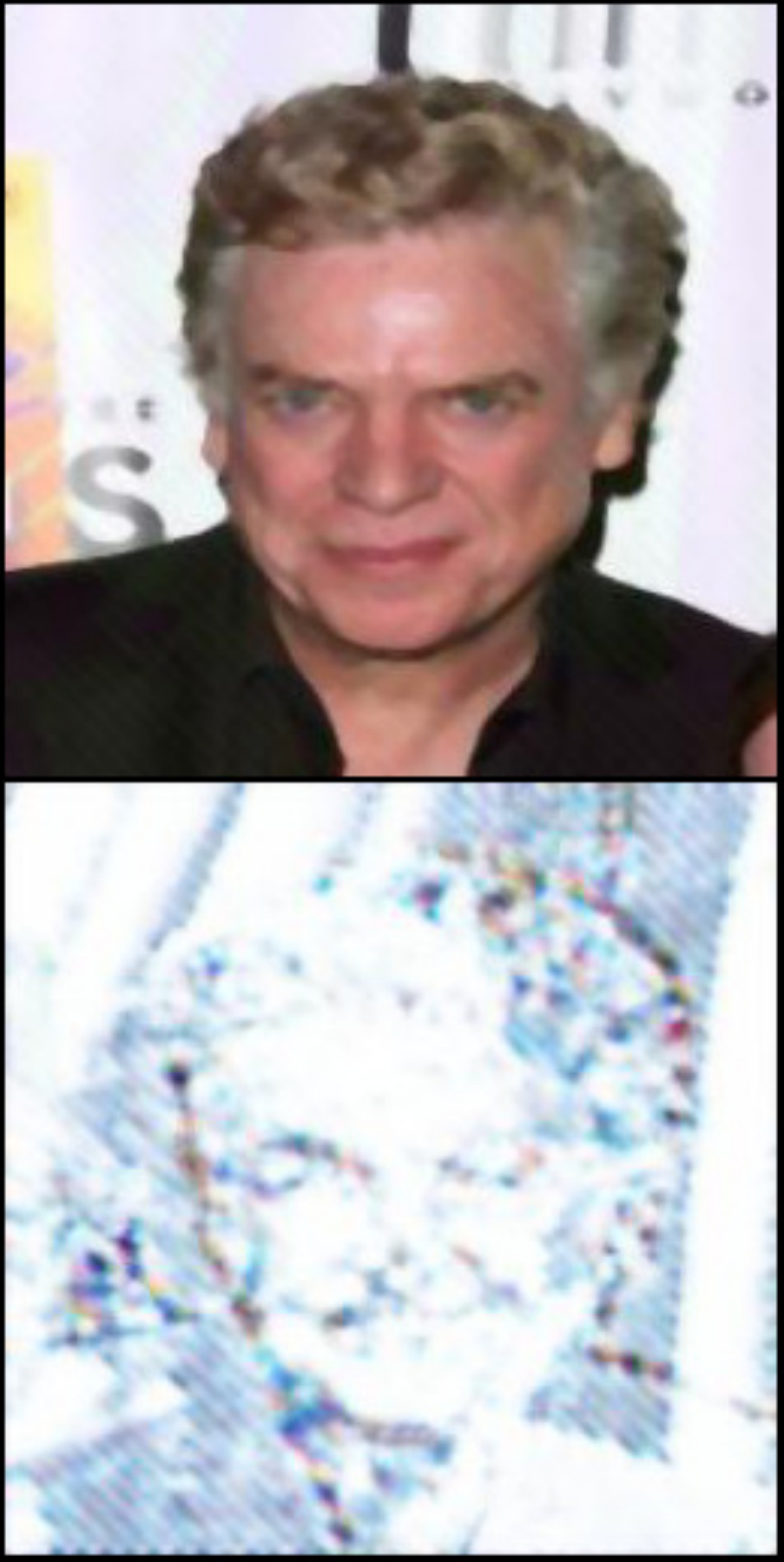}
        &\includegraphics[width=0.12\linewidth]{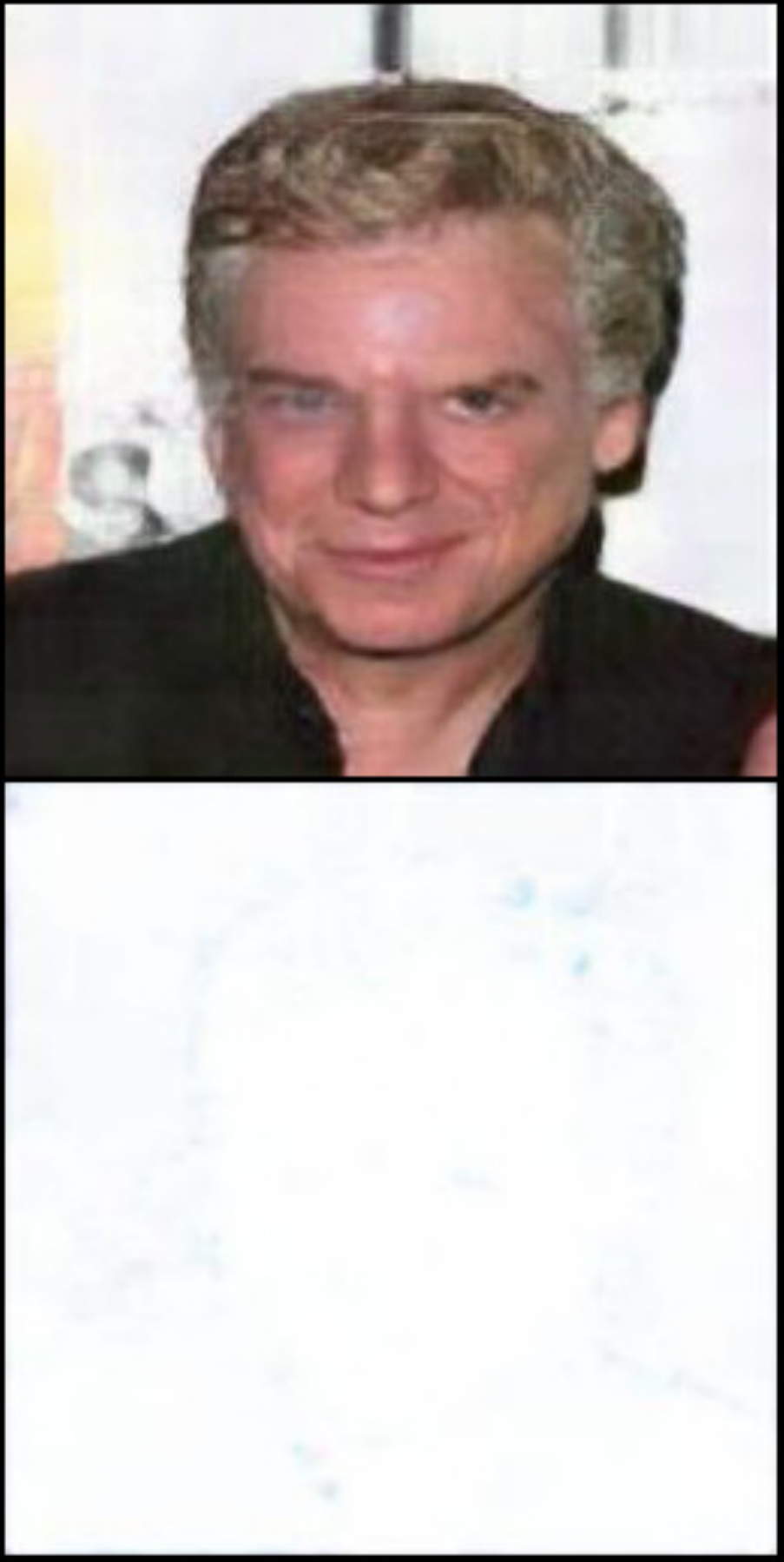}
        &\includegraphics[width=0.12\linewidth]{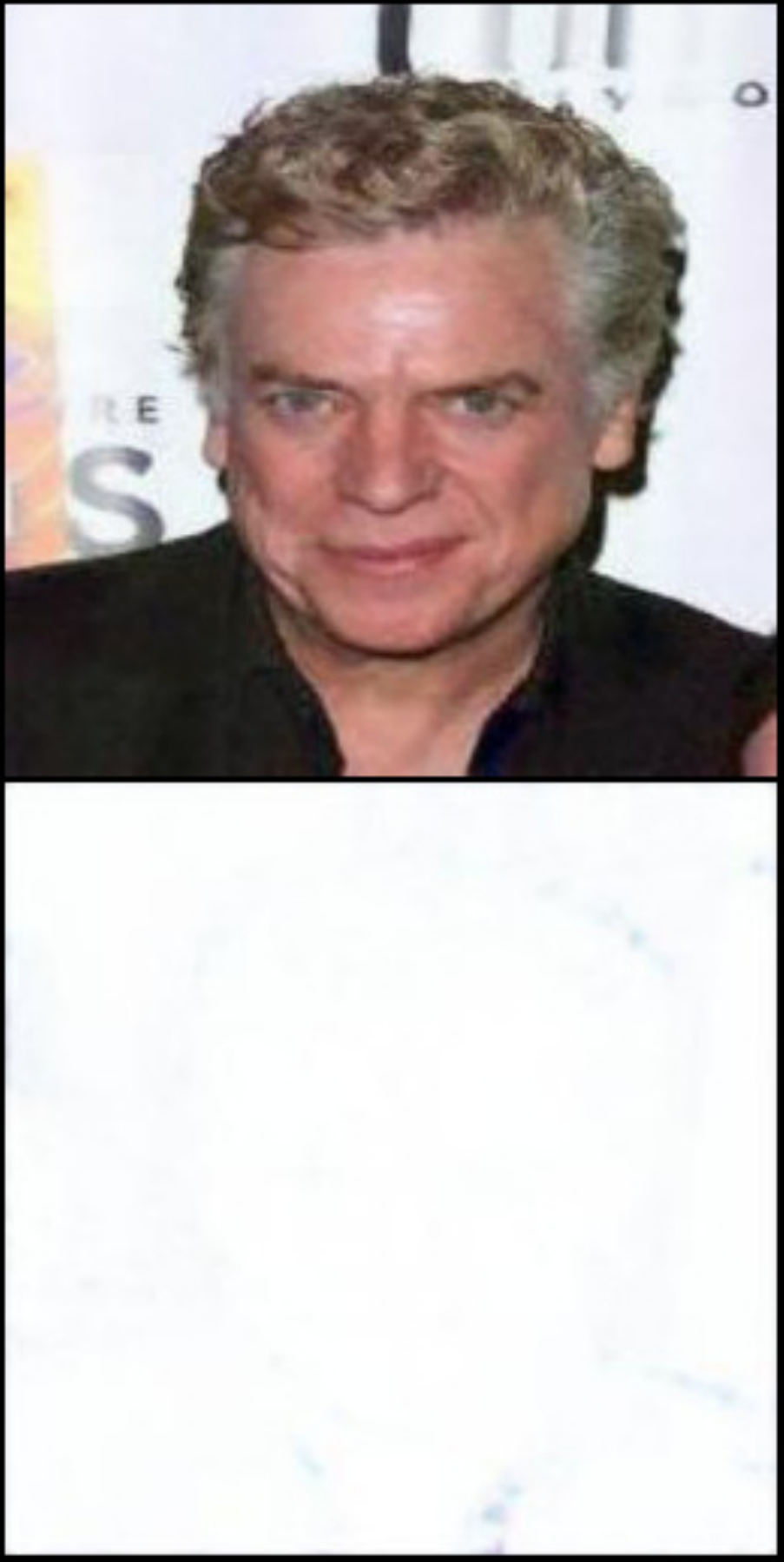} \\
        & &0.584|0.120& 0.455|0.056&0.456|0.155&0.524|0.108&0.245|0.004&0.569|0.005\\
    \end{tabular}%
  }
    \caption{Visualization examples of the attacks on UDH. The images of the first row are containers. The images of the second row are the corresponding revealed secret images after the removal attacks. The two values under each group are VIF-C and VIF-S, respectively.}

  \label{fig:examples_nor}%
\end{figure}

\begin{wraptable}{r}{0.473\linewidth}
  \centering
  \caption{Attack results on UDH}
  \resizebox*{0.47\textwidth}{!}{
  \begin{tabular}{c|cc|cc}
    \toprule
    Attack   & PSNR-C & PSNR-S & VIF-C & VIF-S \\ \hline
GN(0.03) &   30.68     &   15.14     &  0.567     &   0.091    \\
GN(0.05) &   26.36     &   11.16     &  0.346     &   0.043    \\
GB       &   31.3     &    5.35    &    0.523   &   0.098    \\
MB       &   32.74     &   6.07     &   0.579    &  0.058     \\ \hline
PEEL     &   25.44     &   4.57     &   0.302    &  0.004     \\
PEEL-O   &   29.75     &   4.54     &   0.579    &  0.004    \\
    \bottomrule
    \end{tabular}
  }
  \label{tab:normal_vif_psnr}%
\end{wraptable}%
We evaluate the overall performance of our removal attacks and the baselines. In particular, we set the $\delta$ in PEEL-O as 0.05. For GN, we use two different $\delta$ values: 0.03 (GN(0.03)) and 0.05 (GN(0.05)).
We first evaluate the performance of the involved attacks on the three deep hiding schemes. We observe that all the attacks can successfully remove secret images hidden by DS and ISGAN. Different from DS and ISGAN, UDH is more robust in our experiments. Thus, we only show the statistical attack results on UDH (see Supplementary for the attack results on DS and ISGAN). From Table \ref{tab:normal_vif_psnr}, we observe that all baseline attacks have larger PSNR-S and VIF-S values than our PEELs, which indicate our PEELs are superior to attack UDH. We also observe that the PSNR-S of GB is only 5.35 but the corresponding VIF-S is 0.098, much lager than that of GN(0.03). After we check the original results, GB fails in attacking UDH and we confirm again that PSNR is not proper for the quality evaluation of low-quality images and the results of VIF are more accurate. The visualization examples of the attack results on UDH are illustrated in Figure \ref{fig:examples_nor}.

\begin{figure}[h]
  \centering
  \resizebox*{\textwidth}{!}{
    \begin{tabular}{cccccccc}
        & &GN(0.03)  &  GN(0.05) &GB&MB&  PEEL & PEEL-O \\
        \rotatebox{90}{$\ \ \ \ \ \ \ \ \ \ \ \  $UDH-AE}& \includegraphics[width=0.12\linewidth]{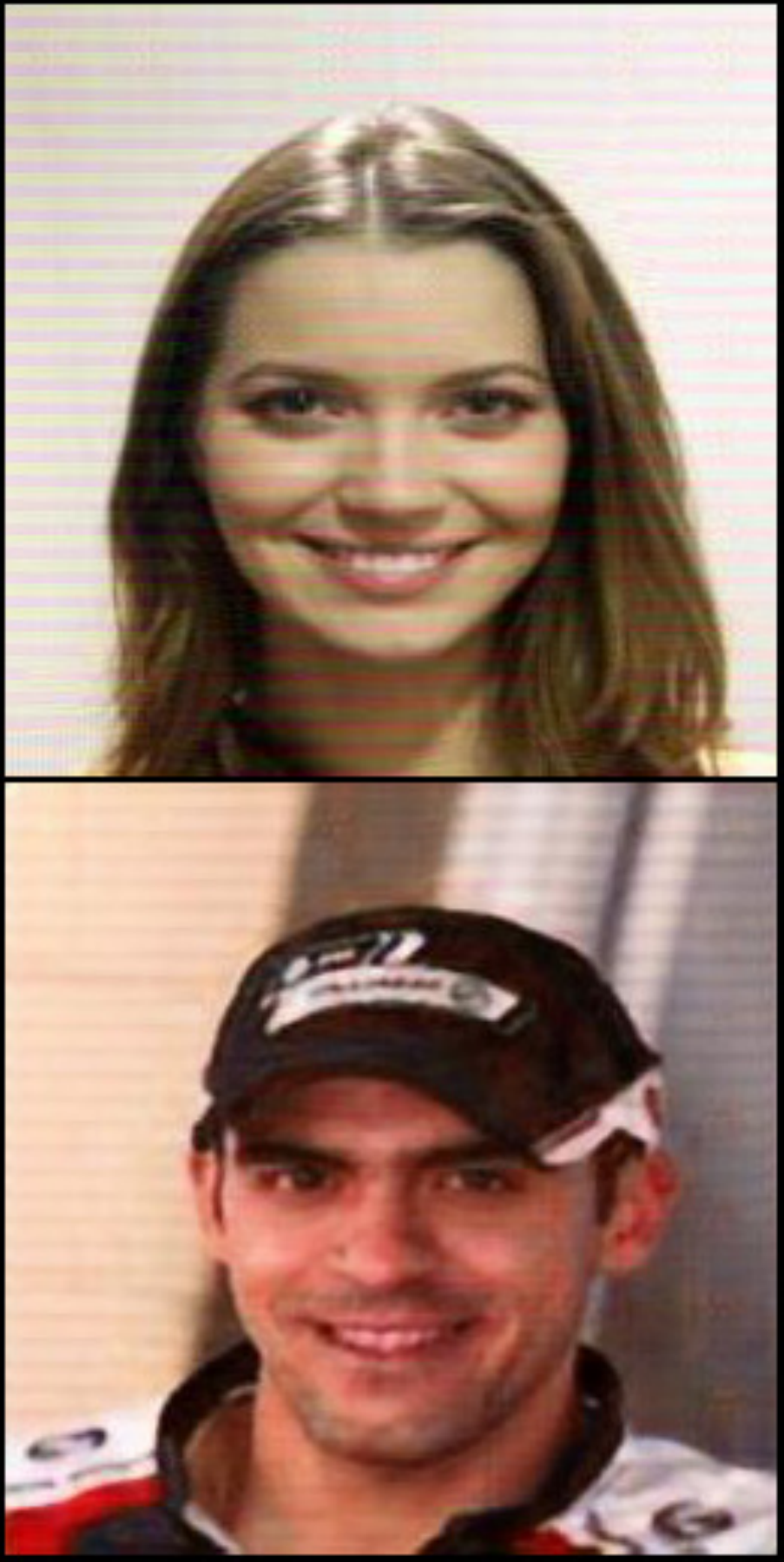}
        &\includegraphics[width=0.12\linewidth]{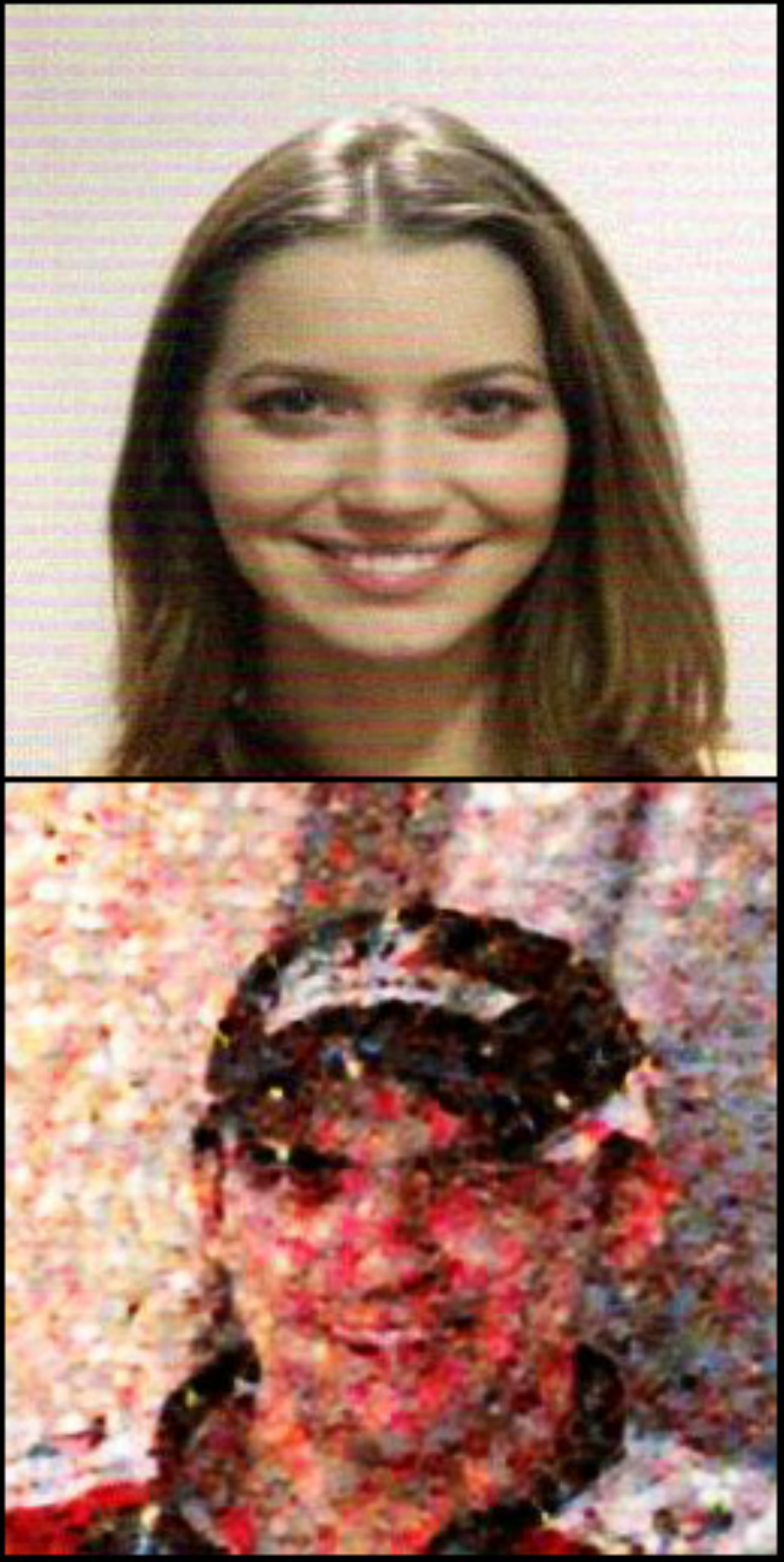}
        &\includegraphics[width=0.12\linewidth]{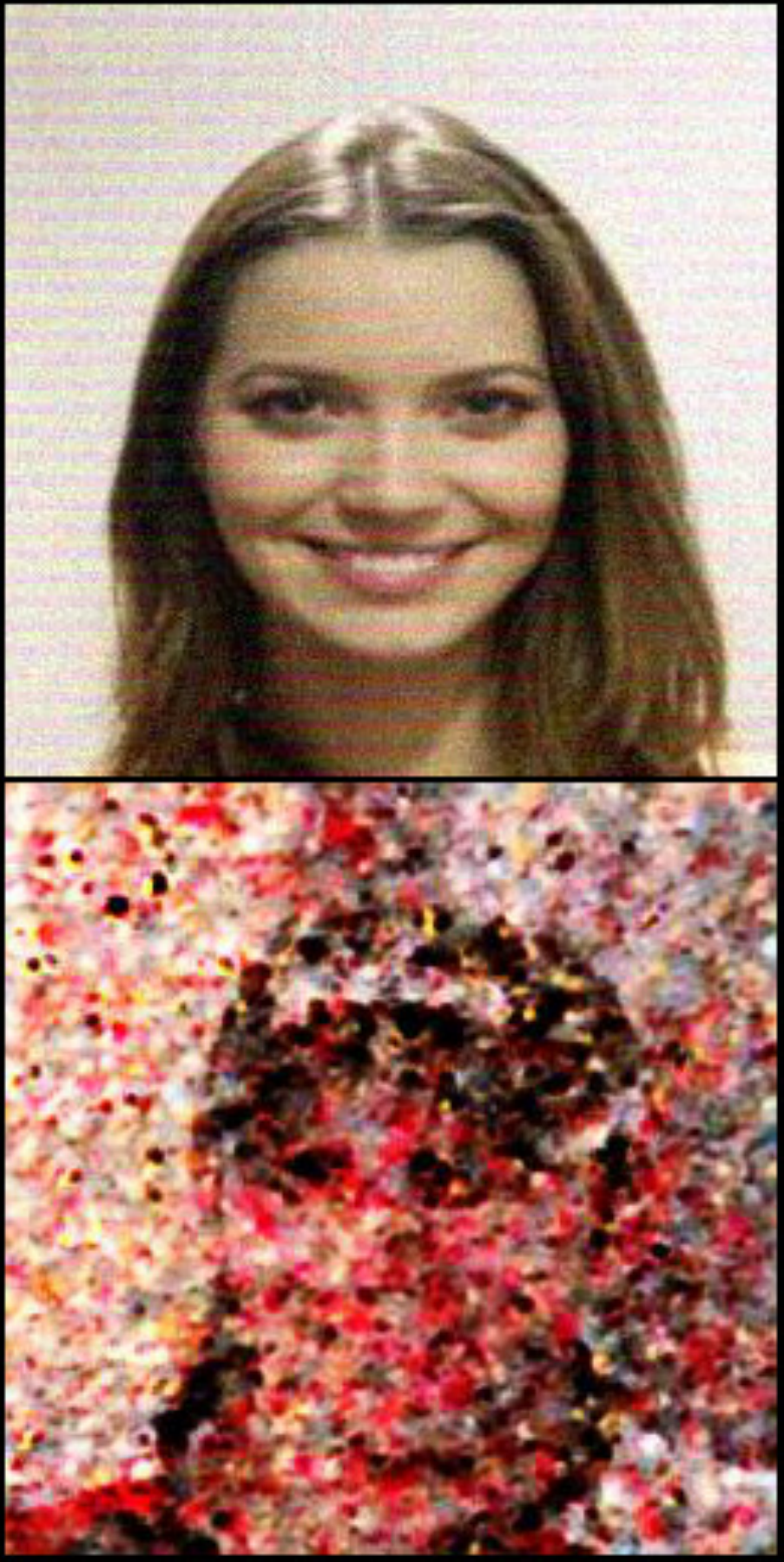}
        &\includegraphics[width=0.12\linewidth]{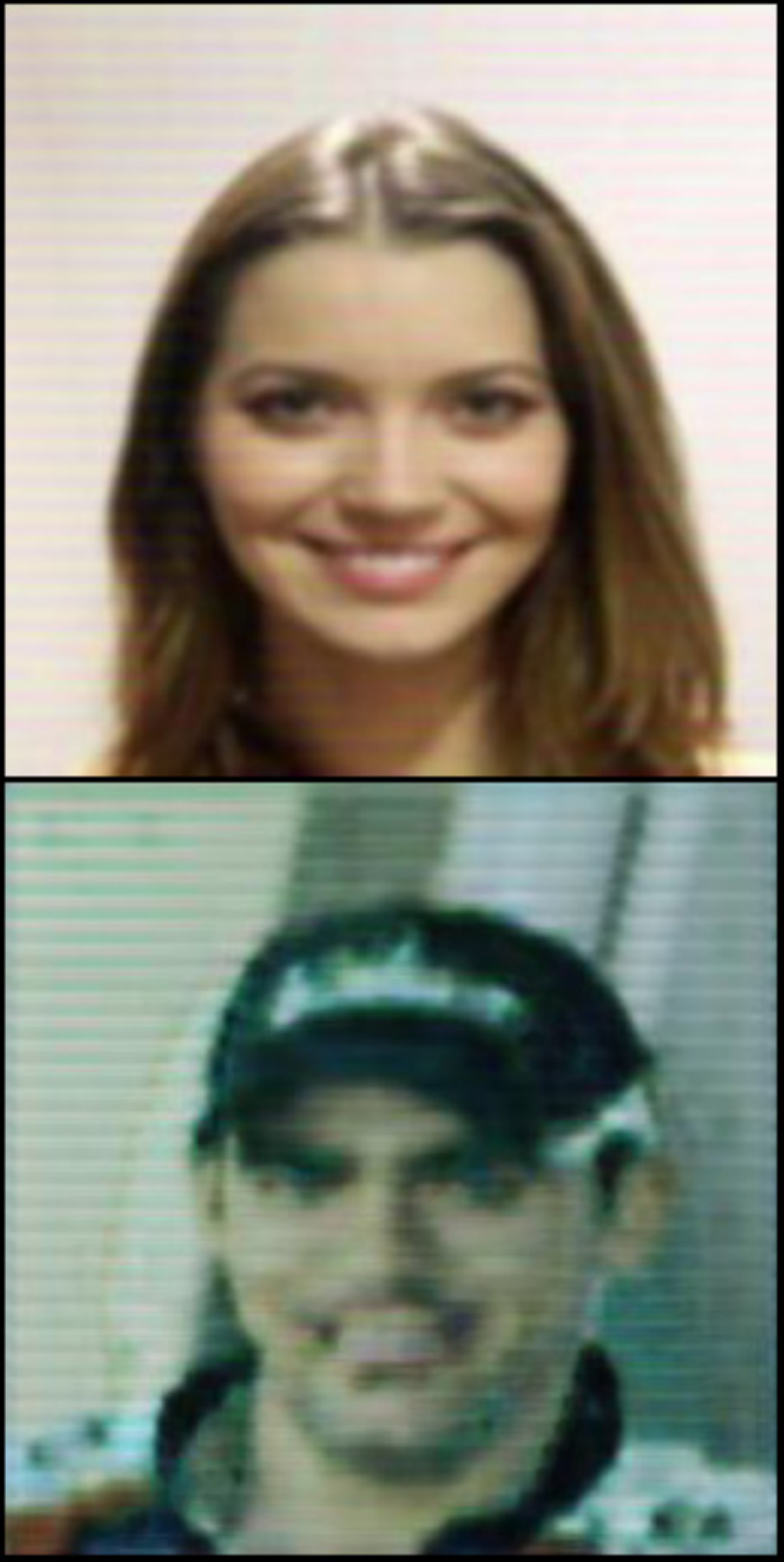}
        &\includegraphics[width=0.12\linewidth]{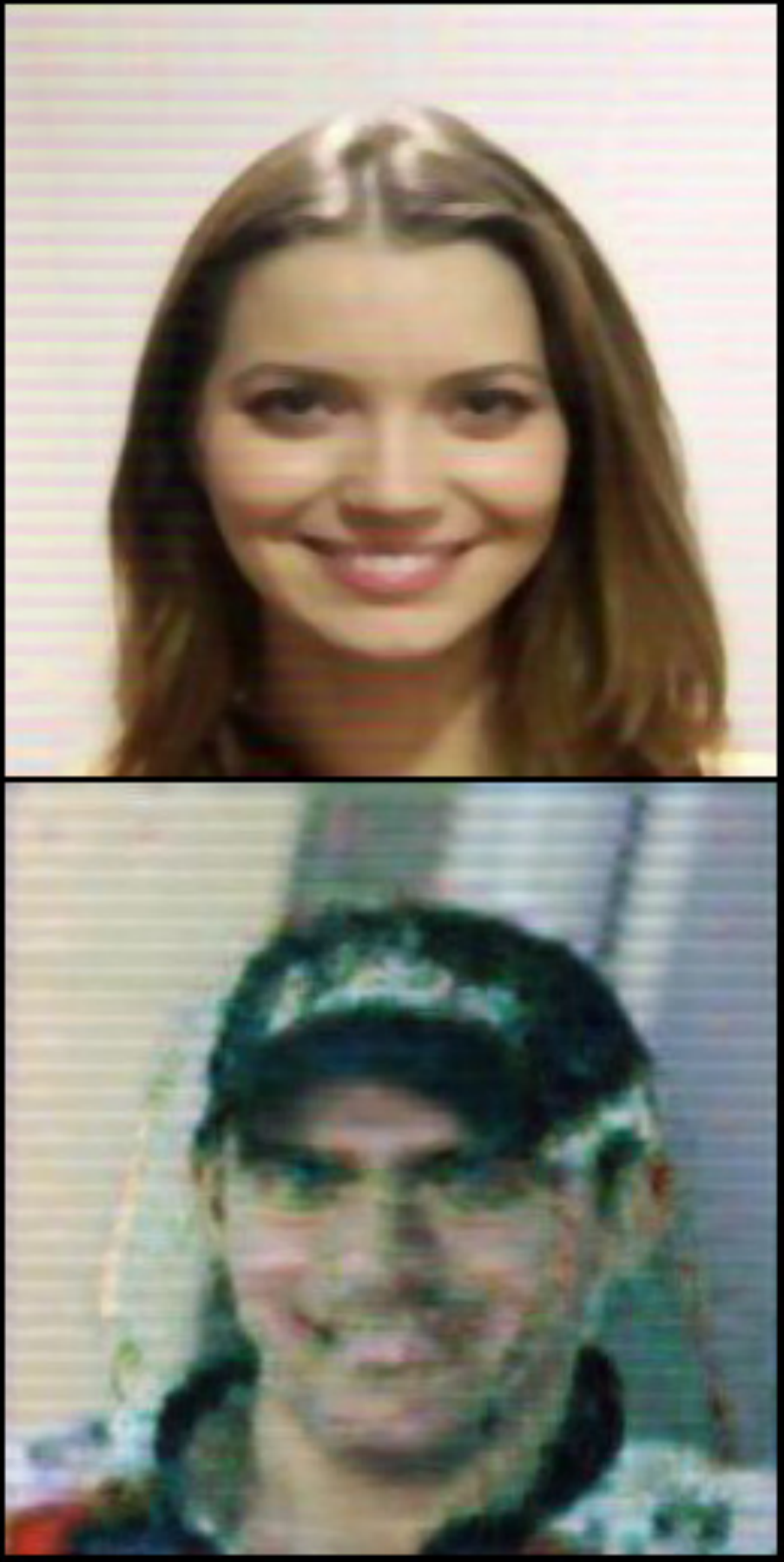}
        &\includegraphics[width=0.12\linewidth]{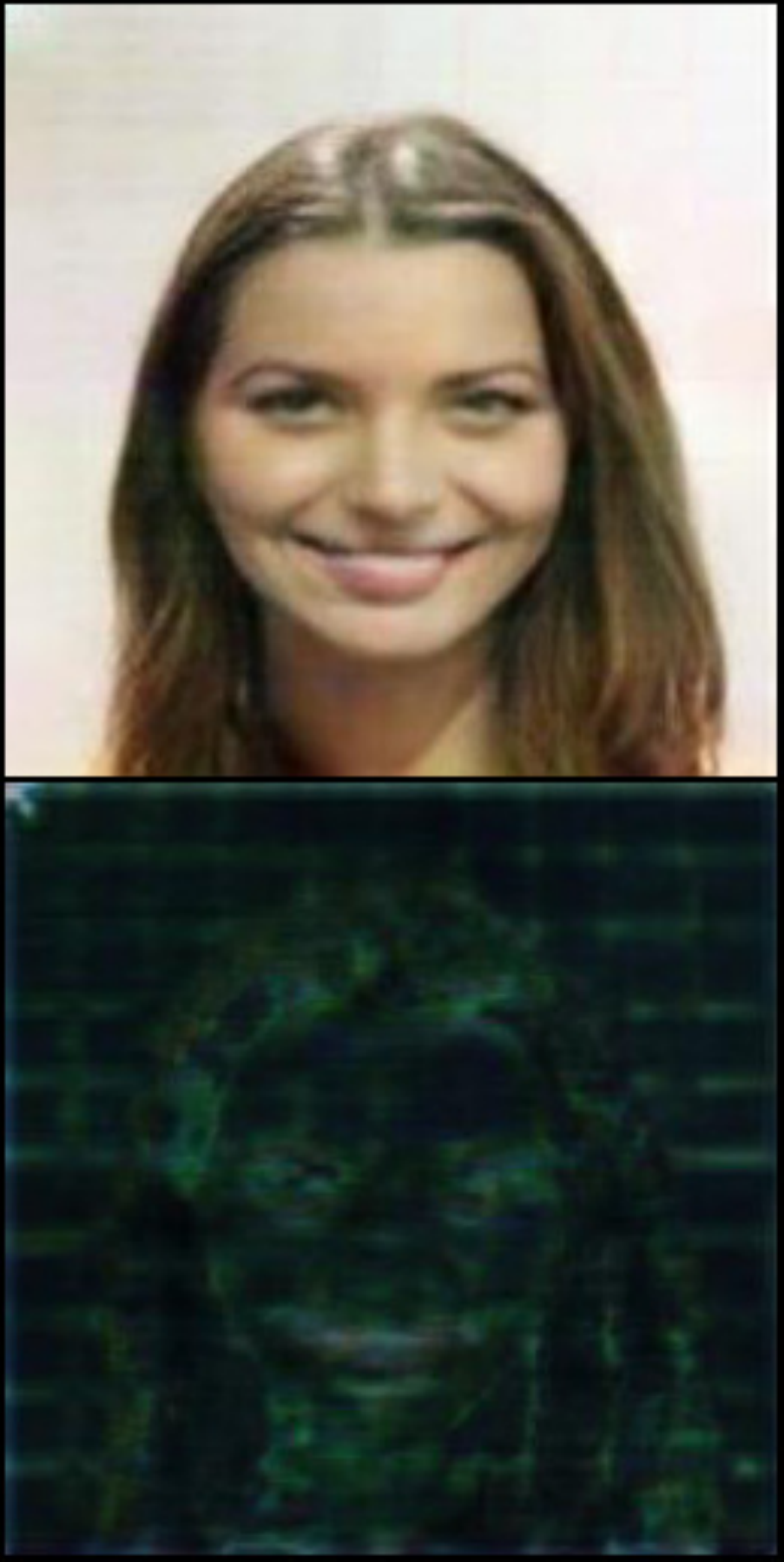}
        &\includegraphics[width=0.12\linewidth]{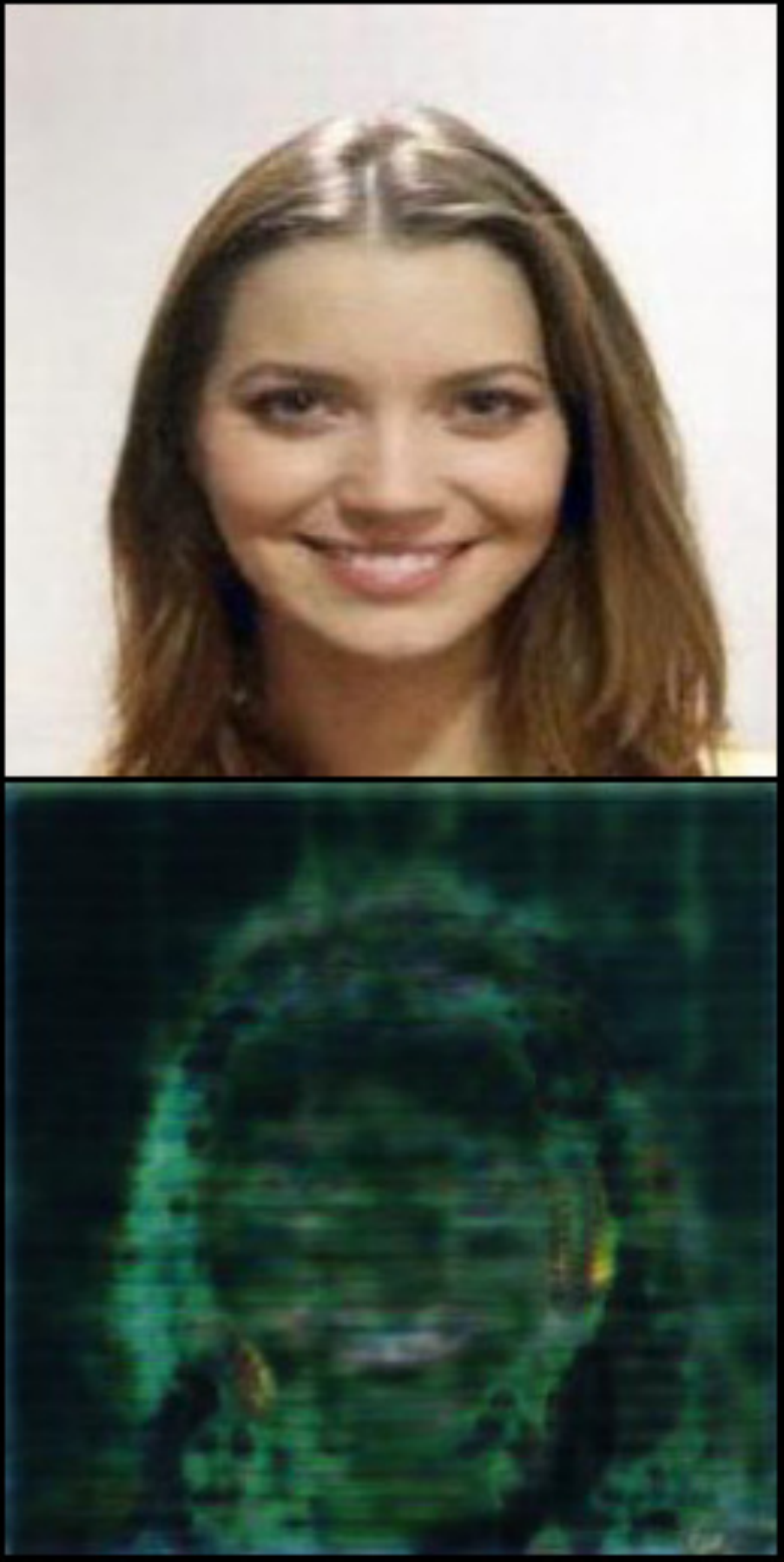} \\
        & &0.575|0.161& 0.448|0.076&0.522|0.255&0.608|0.265&0.363|0.018&0.559|0.042\\

        \rotatebox{90}{$\ \ \ \ \ \ \ \ \ \ \ \ \ \  $DS-AE}& \includegraphics[width=0.12\linewidth]{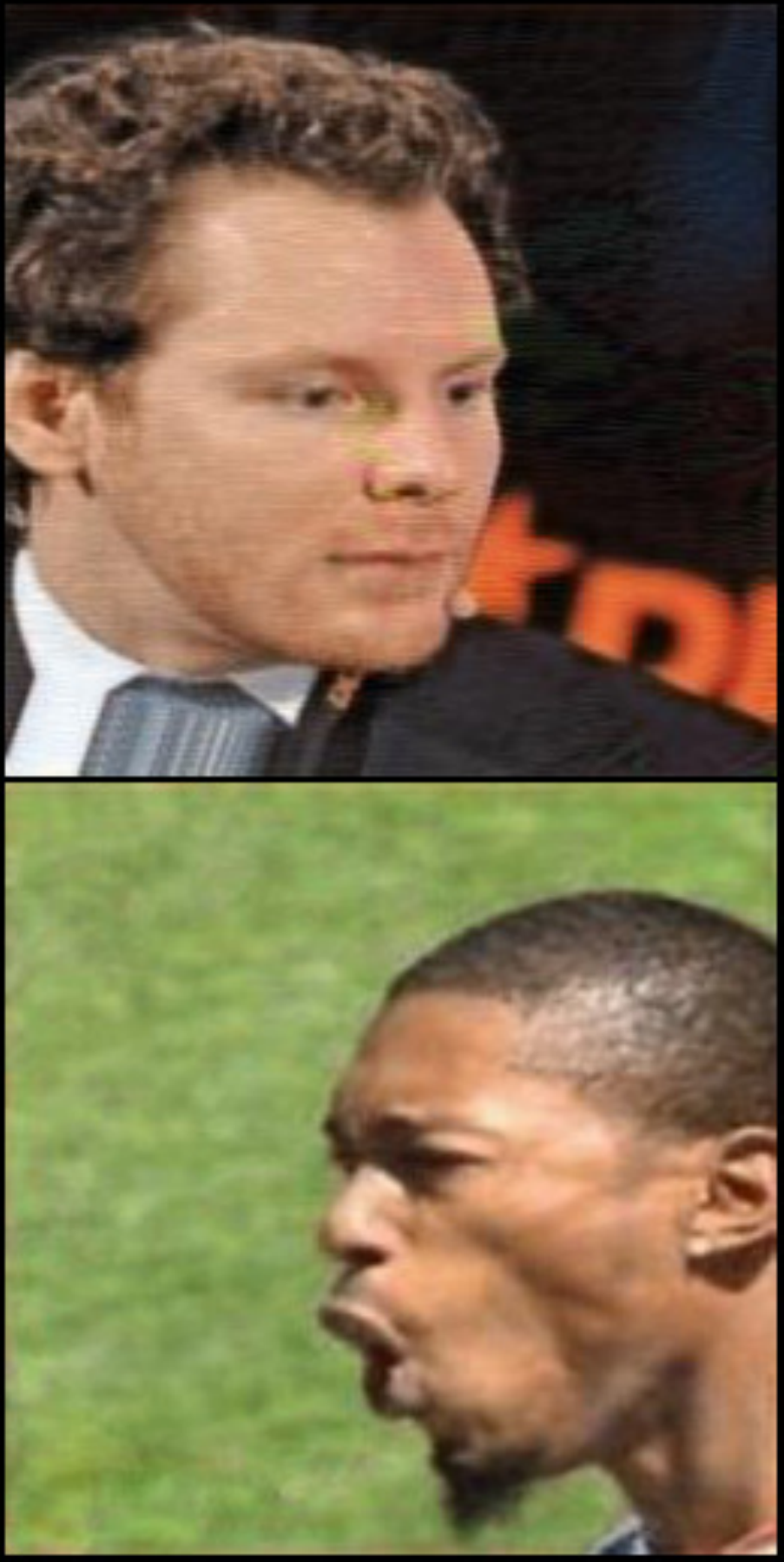}
        & \includegraphics[width=0.12\linewidth]{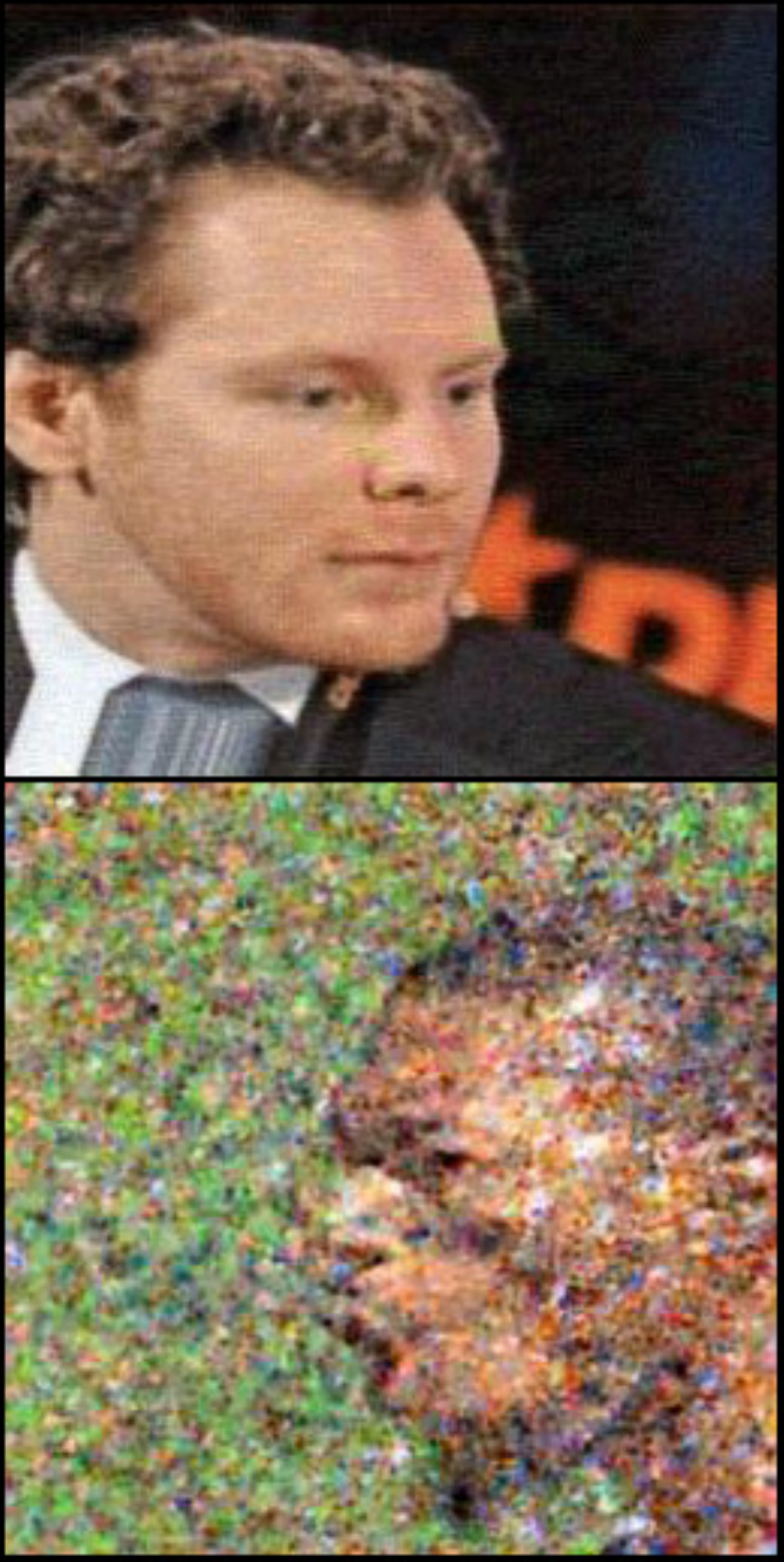}
        &   \includegraphics[width=0.12\linewidth]{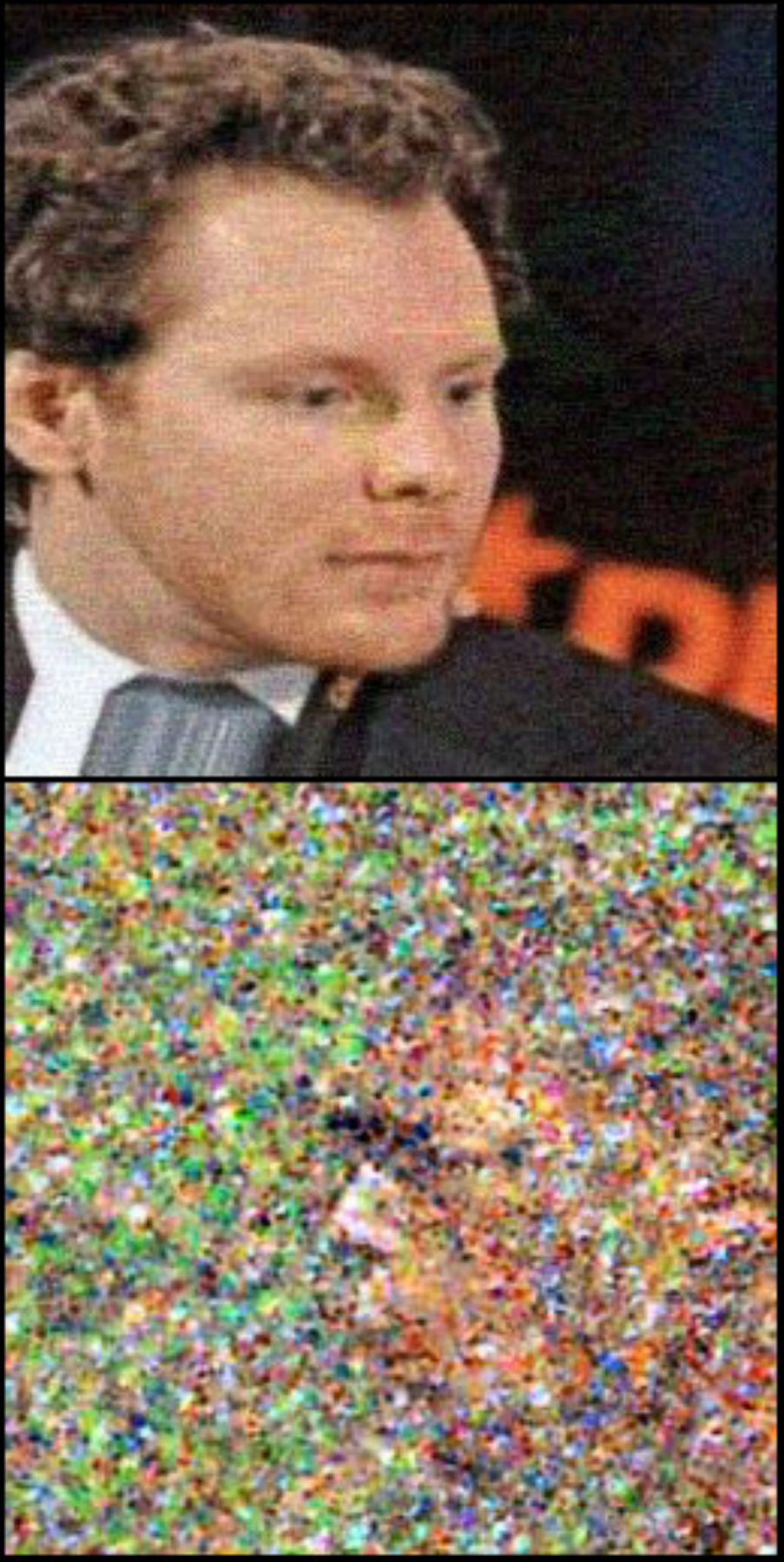}
        & \includegraphics[width=0.12\linewidth]{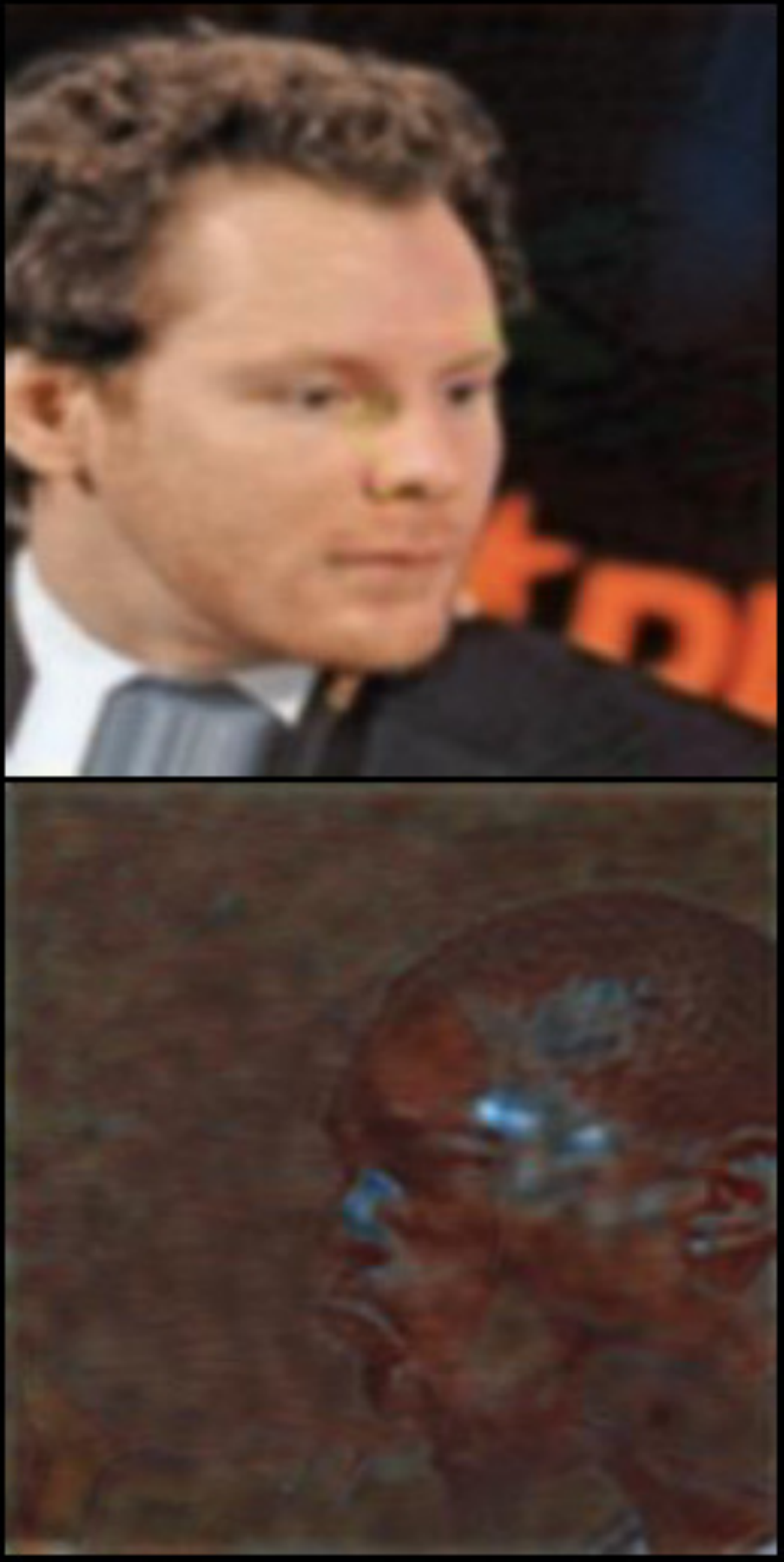}
        & \includegraphics[width=0.12\linewidth]{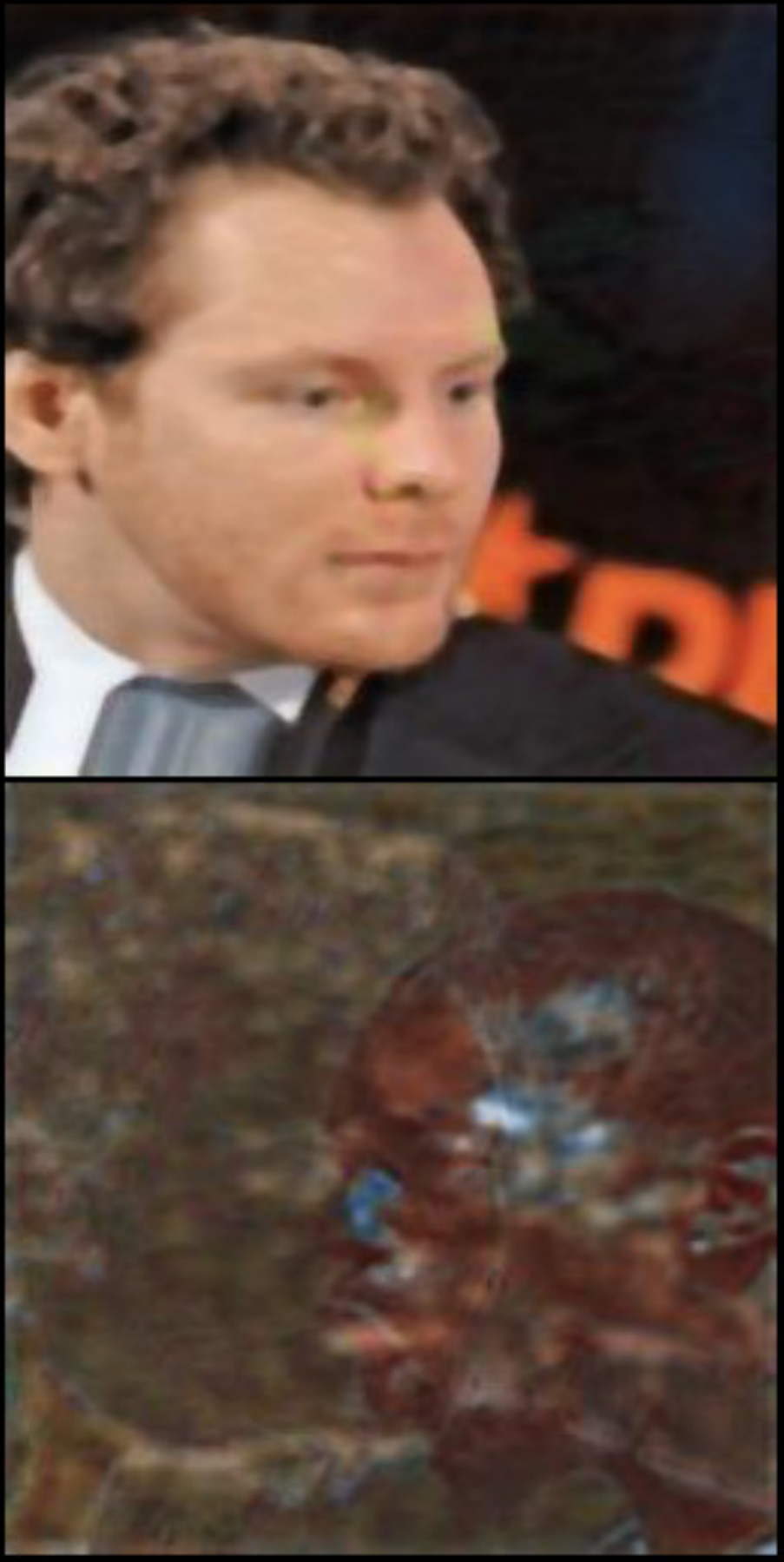}
        & \includegraphics[width=0.12\linewidth]{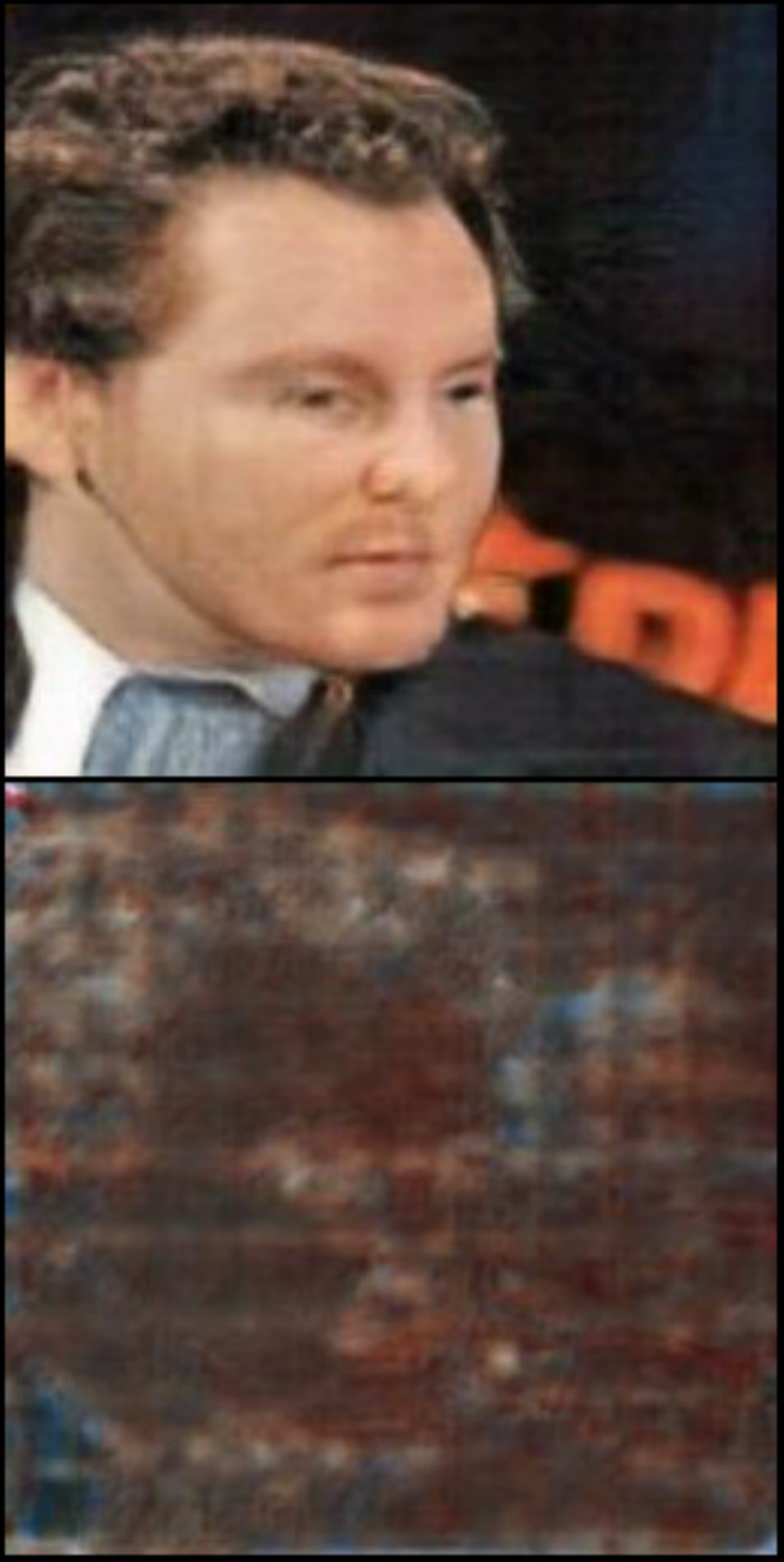}
        & \includegraphics[width=0.12\linewidth]{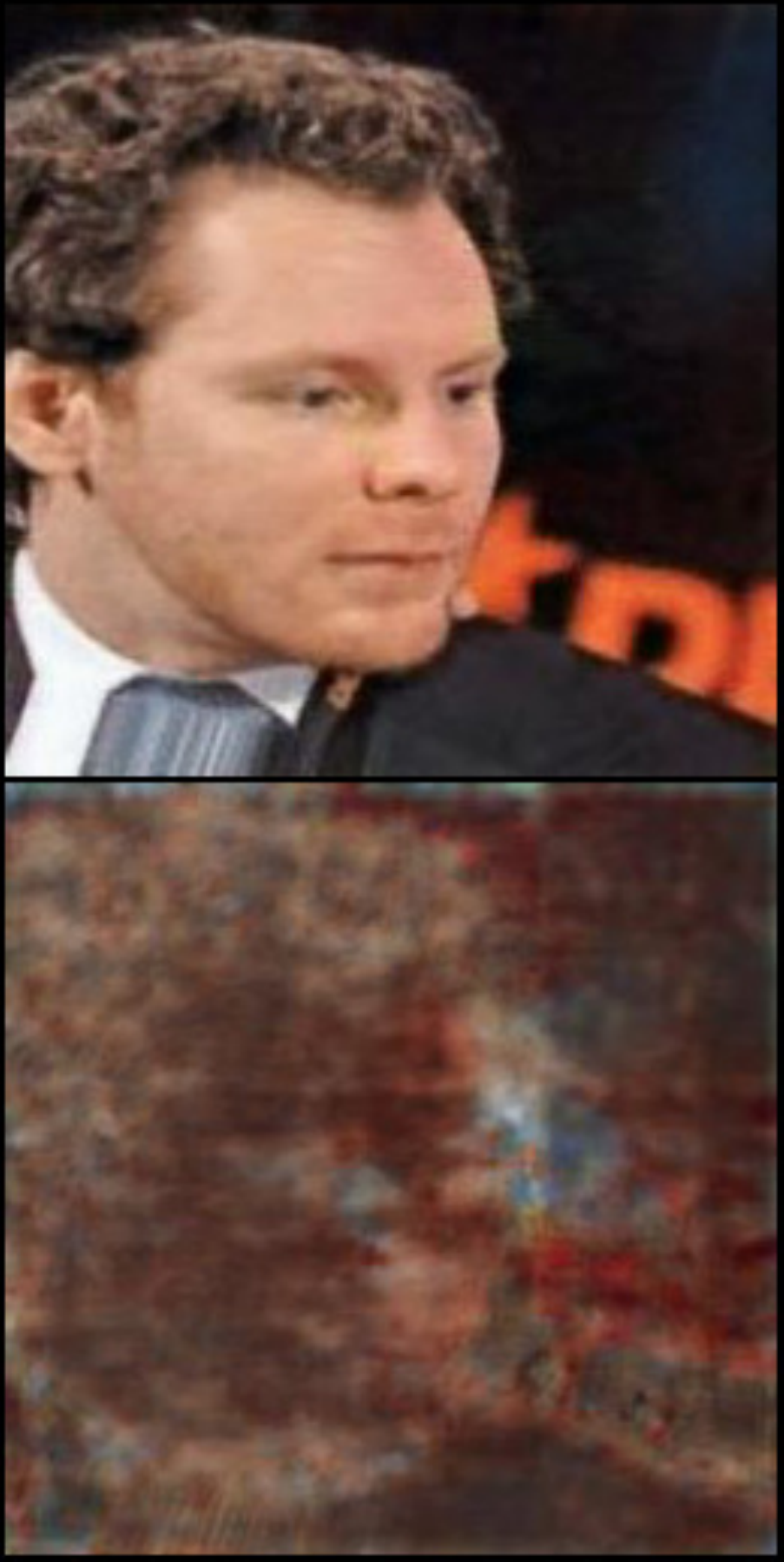} \\
        & & 0.558|0.081& 0.424|0.042&0.472|0.127&0.508|0.104&0.256|0.028& 0.558|0.056\\

        \rotatebox{90}{$\ \ \ \ \ \ \ \ \ \  $ISGAN-AE}& \includegraphics[width=0.12\linewidth]{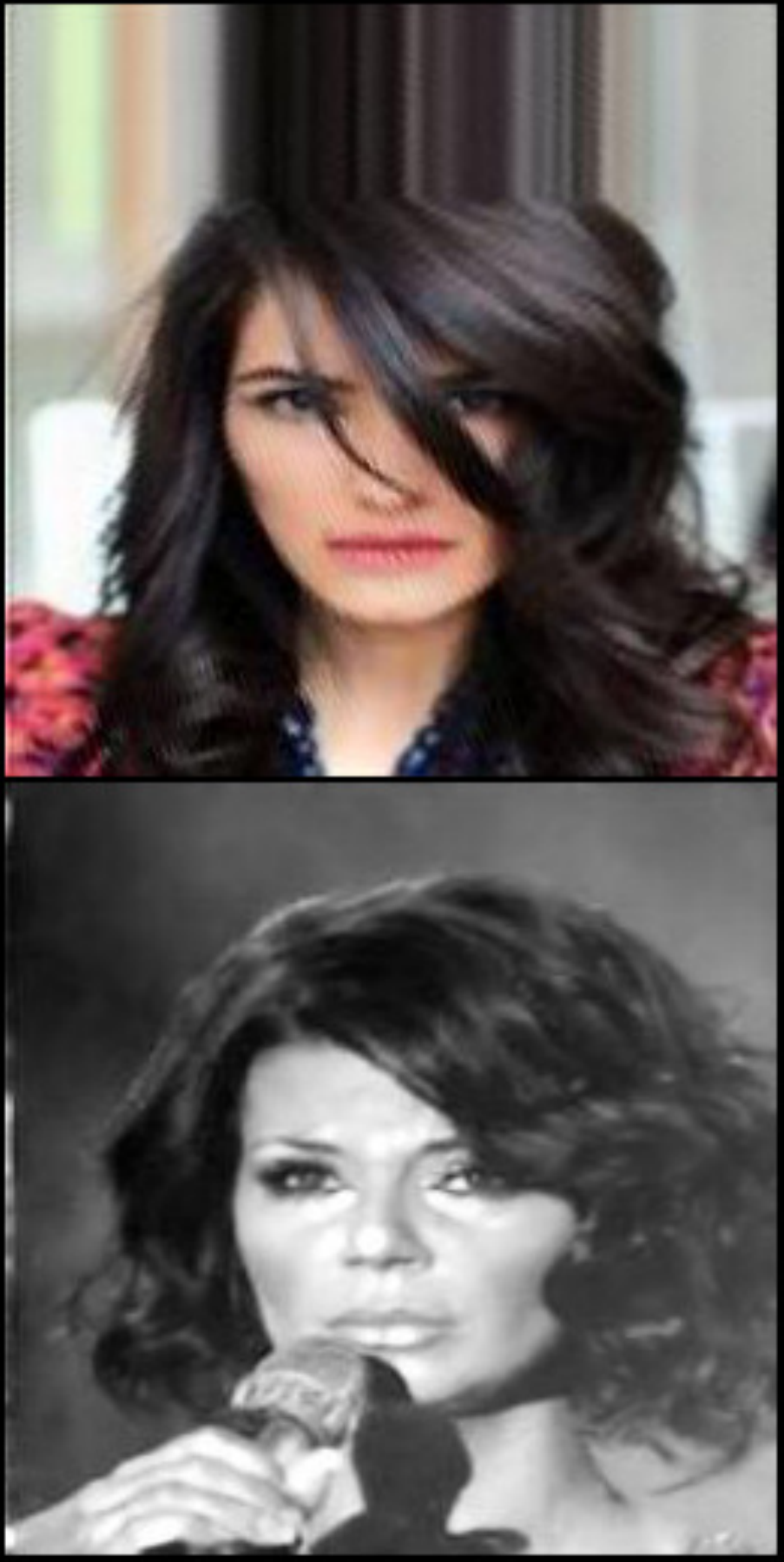}
        & \includegraphics[width=0.12\linewidth]{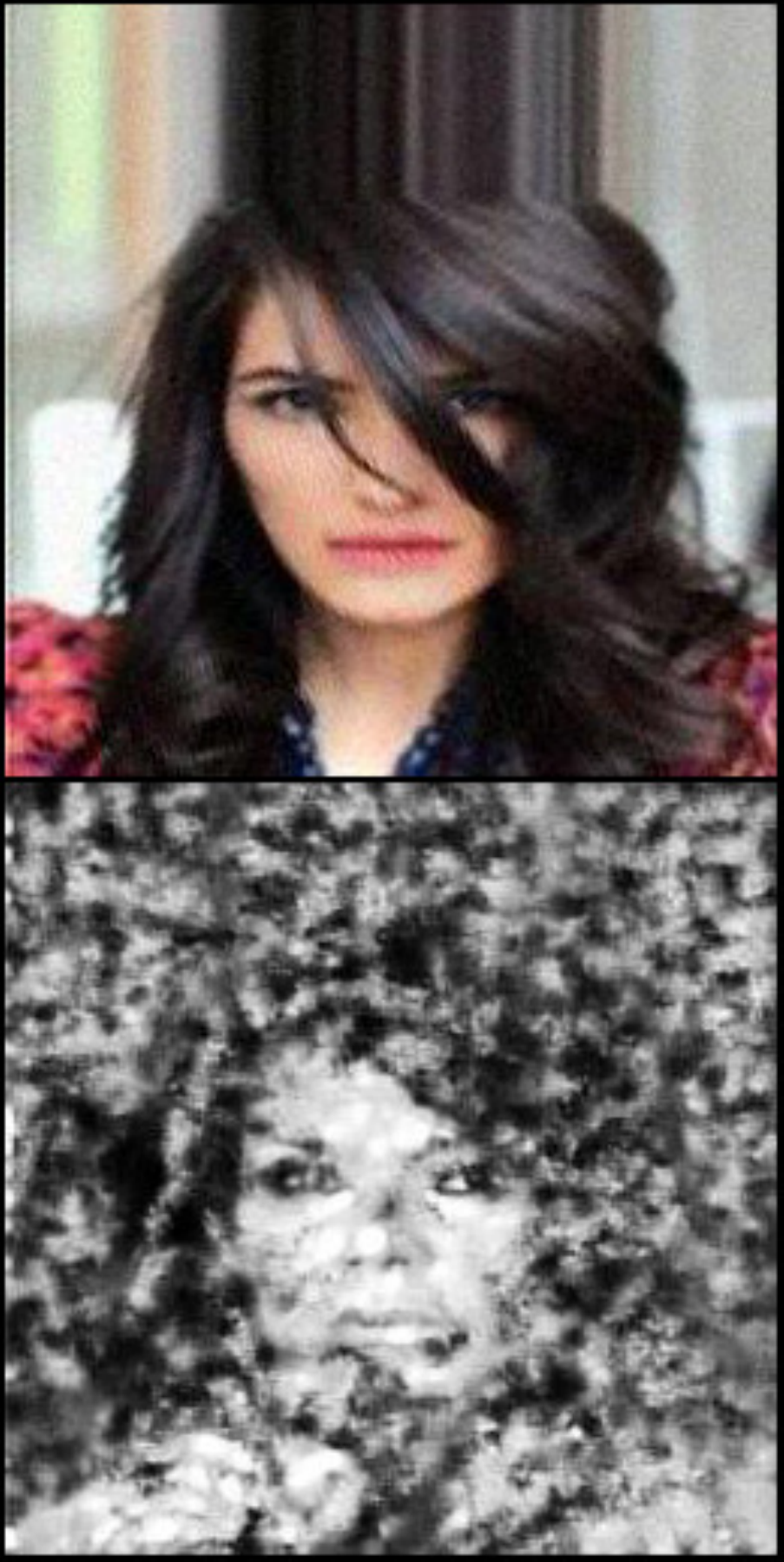}
        &   \includegraphics[width=0.12\linewidth]{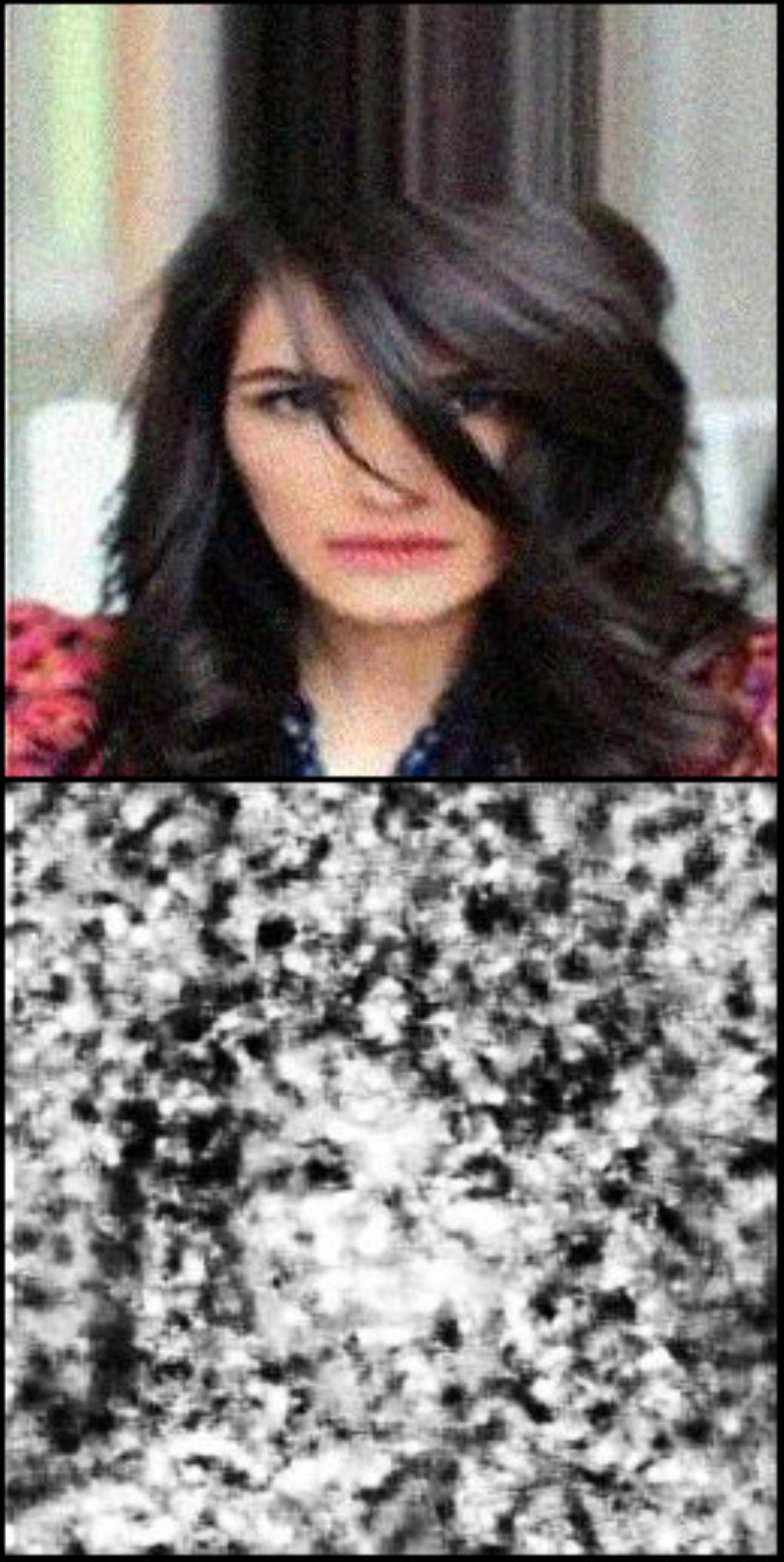}
        & \includegraphics[width=0.12\linewidth]{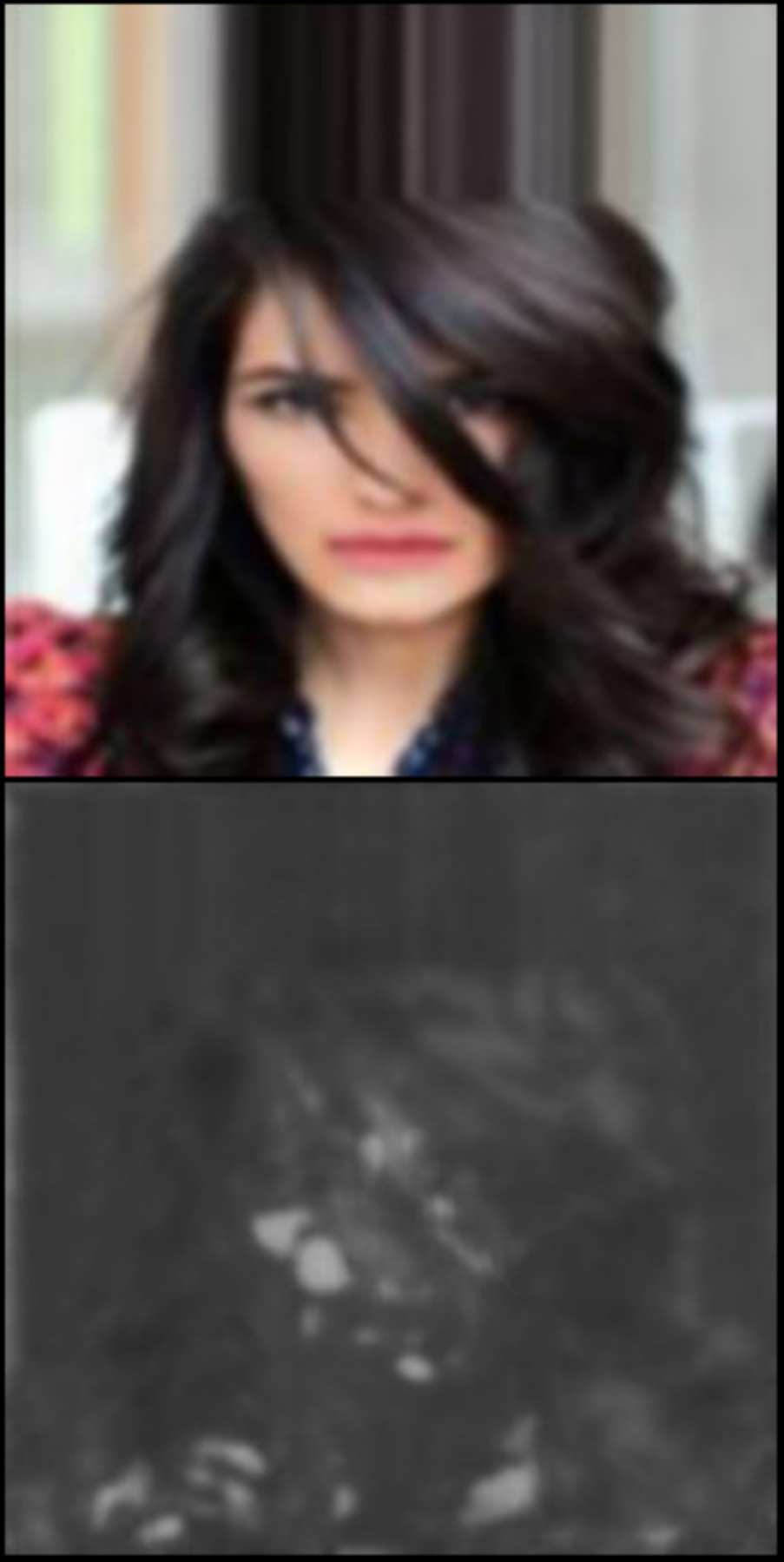}
        & \includegraphics[width=0.12\linewidth]{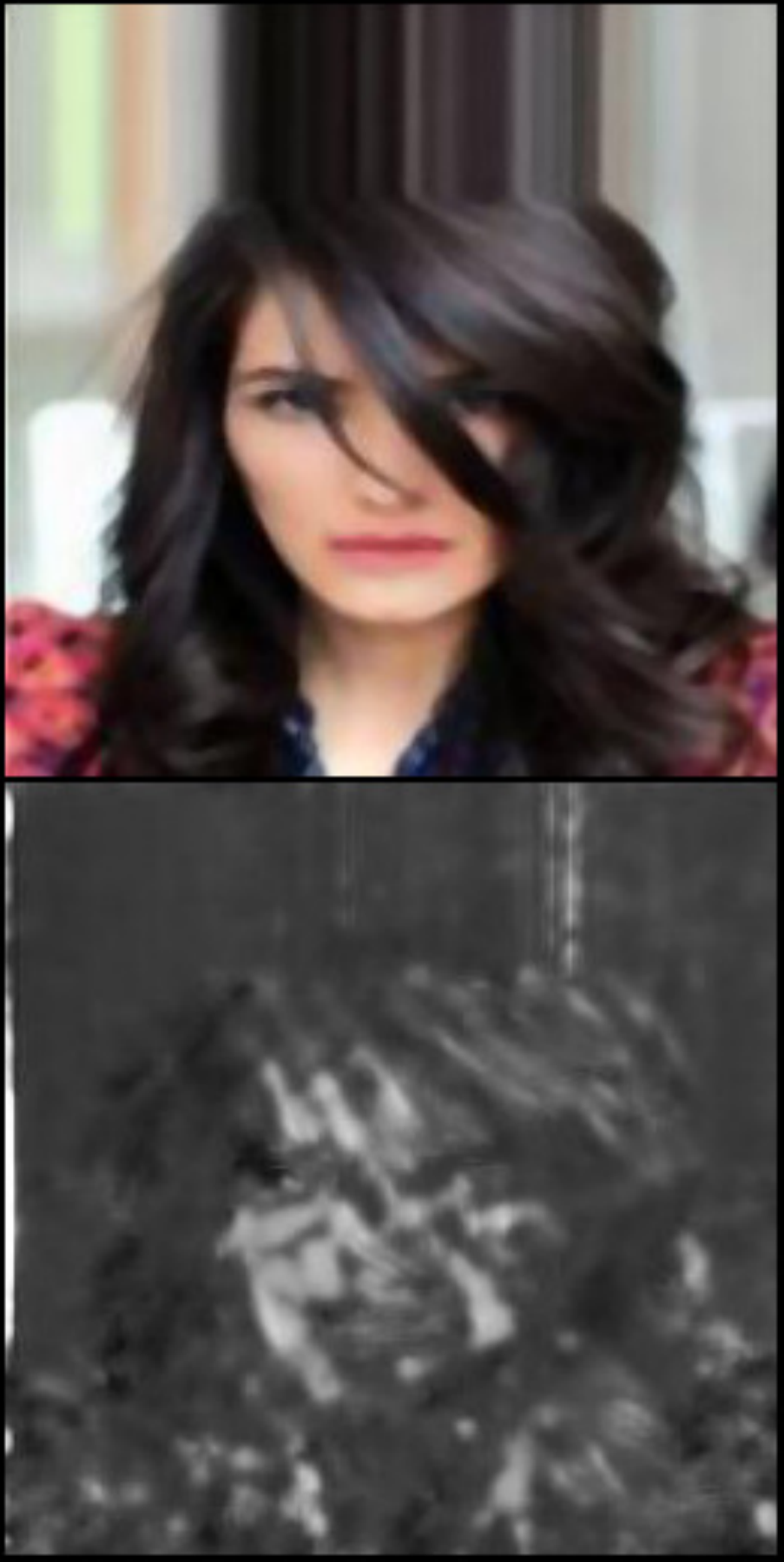}
        & \includegraphics[width=0.12\linewidth]{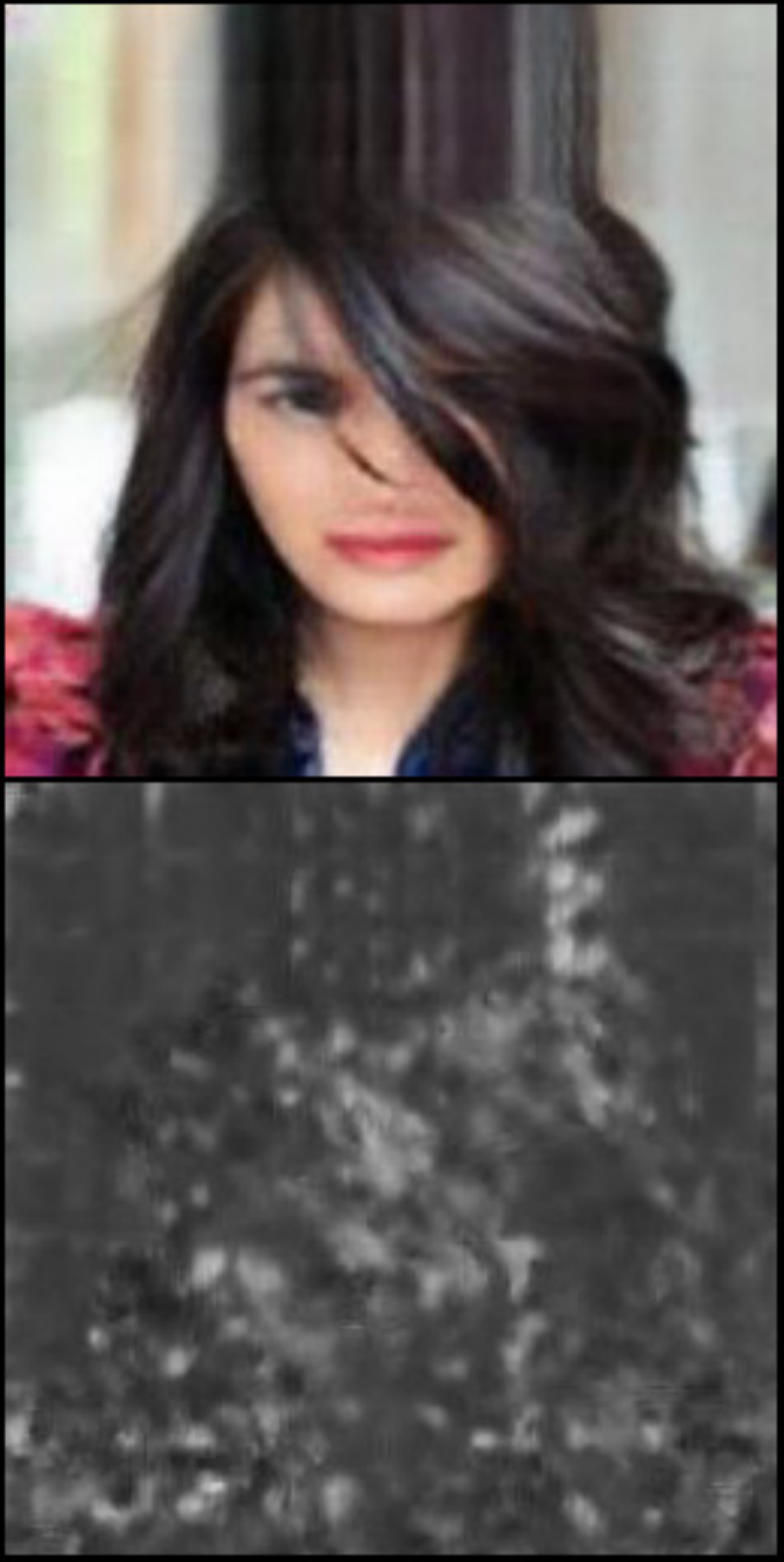}
        & \includegraphics[width=0.12\linewidth]{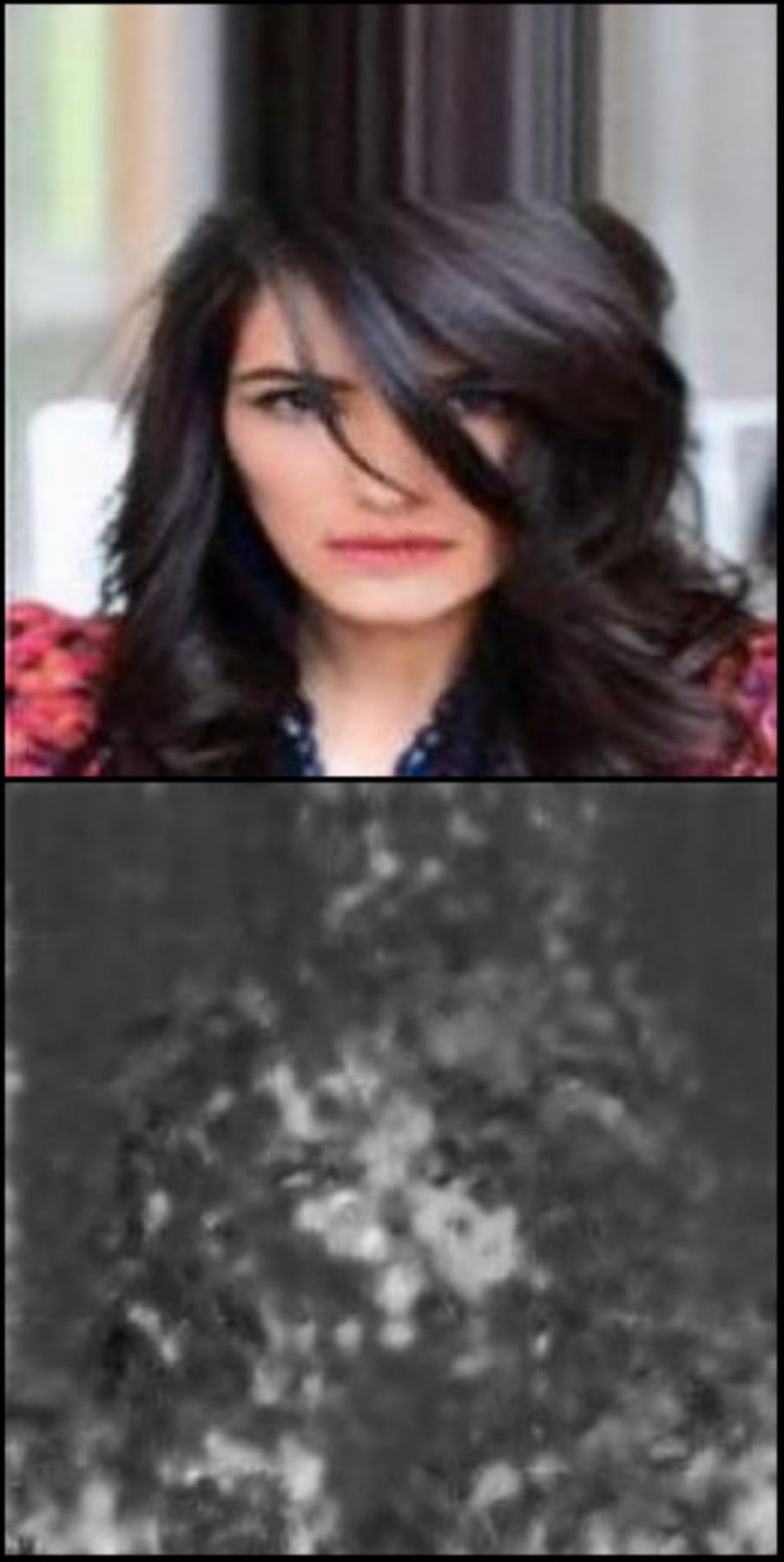} \\
        & & 0.581|0.092& 0.448|0.049&0.607|0.049&0.665|0.051& 0.372|0.024& 0.679|0.053\\
    \end{tabular}%
  }
    \caption{Visualization examples of the attacks on the adversarially trained deep hiding schemes. The two values under each group are VIF-C and VIF-S, respectively.}
  \label{fig:examples_adv}%
\end{figure}

Next, we evaluate these attacks on the adversarially trained deep hiding schemes. The experimental results are illustrated in Table \ref{tab:adv_vif_psnr}. We observe that our PEELs still achieve lower VIF-S values on all the enhanced schemes.
On the contrary, the baselines fail in attacking the enhanced versions in most cases since the corresponding VIF-S values in Table \ref{tab:adv_vif_psnr} are much larger than those of PEELs. For example, the VIF-S value of GB on DS-AE is 0.159, much larger than those of PEELs (0.33 and 0.66).
In addition, we can observe that PSNR-C and VIF-C values of PEEL-O are much higher than those of PEEL, which shows the effectiveness of the quality optimization strategy.
In order to visually exhibit the experimental results, we show some examples in Figure \ref{fig:examples_adv}. In Figure \ref{fig:examples_adv}, we can acquire obvious visual content of secret images from the attack results when we use the baselines to attack UDH-AE. As for attacking other adversarially trained schemes, the visual content of the secret images is still reserved in some cases. On the contrary, our PEELs can remove the secret images completely. More visualization examples can be found in Supplementary.


\begin{table}[t]
  \centering
  \caption{Attack results on the adversarially trained deep hiding schemes}
  \resizebox*{0.95\textwidth}{!}{
  \begin{tabular}{c|ccc|ccc}
    \toprule
    \multirow{2}{*}{Attack} & \multicolumn{3}{c|}{PSNR(PSNR-C|PSNR-S)} & \multicolumn{3}{c}{VIF(VIF-C|VIF-S)} \\
\cline{2-7}          & UDH-AE & DS-AE  & ISGAN-AE & UDH-AE & DS-AE & ISGAN-AE \\
\hline
    GN(0.03) & 30.68|19.33 & 30.58|15.59 & 30.63|13.39 & 0.564|0.160 & 0.564|0.105 & 0.584|0.084 \\
    GN(0.05) & 26.35|14.28 & 26.27|11.84 & 26.29|9.93 & 0.432|0.072 & 0.432|0.049 & 0.454|0.040 \\
    GB&30.65|15.43&32.65|11.72&32.42|10.49  &0.486|0.240&0.532|0.159&0.568|0.042\\
    MB&32.05|17.11&34.09|12.88&34.4|11.12  &0.548|0.246&0.582|0.138&0.627|0.051\\
    \hline
    PEEL  & 23.91|7.78 & 26.09|11.40 & 26.56|10.56 & 0.276|0.019 & 0.313|0.033 & 0.332|0.022 \\
    PEEL-O & 29.03|8.56 & 30.19|12.12 & 29.29|11.42 & 0.543|0.040 & 0.592|0.066 & 0.651|0.048 \\
    \bottomrule
    \end{tabular}%
  }
  \label{tab:adv_vif_psnr}%
\end{table}%

\begin{wraptable}{r}{0.63\linewidth}
  \centering
    \caption{Attack results of PEELs with different $\delta$}
    \resizebox*{0.63 \textwidth}{!}{
    \begin{tabular}{c|cc|cc}
      \toprule
      \multirow{2}{*}{Attack} & \multicolumn{2}{c|}{PSNR} & \multicolumn{2}{c}{VIF} \\
  \cline{2-5}          & UDH-C & UDH-P & UDH-C & UDH-P \\
      \hline
      PEEL  &  23.88|10.51     & 24.43|12.44 &   0.260|0.014    & 0.293|0.038 \\
      PEEL-O(0.05) & 27.31|10.67 & 29.43|16.81 & 0.500|0.015 & 0.569|0.157 \\
      PEEL-O(0.1) & -     & 28.17|14.54 & -     & 0.480|0.082 \\
      PEEL-O(0.2) & -     & 26.92|11.97 & -     & 0.402|0.036 \\
      \bottomrule
      \end{tabular}%
    \label{tab:udh_adv}%
    }
\end{wraptable}

\subsection{Impact of Adversarial Training}\label{sec:exp:adv}
\begin{figure}
  \centering
  \resizebox*{0.6\textwidth}{!}{
   \begin{tabular}{cccc}
        PEEL  &  PEEL-O(0.05)&PEEL-O(0.1)& PEEL-O(0.2) \\
        \includegraphics[width=0.12\linewidth]{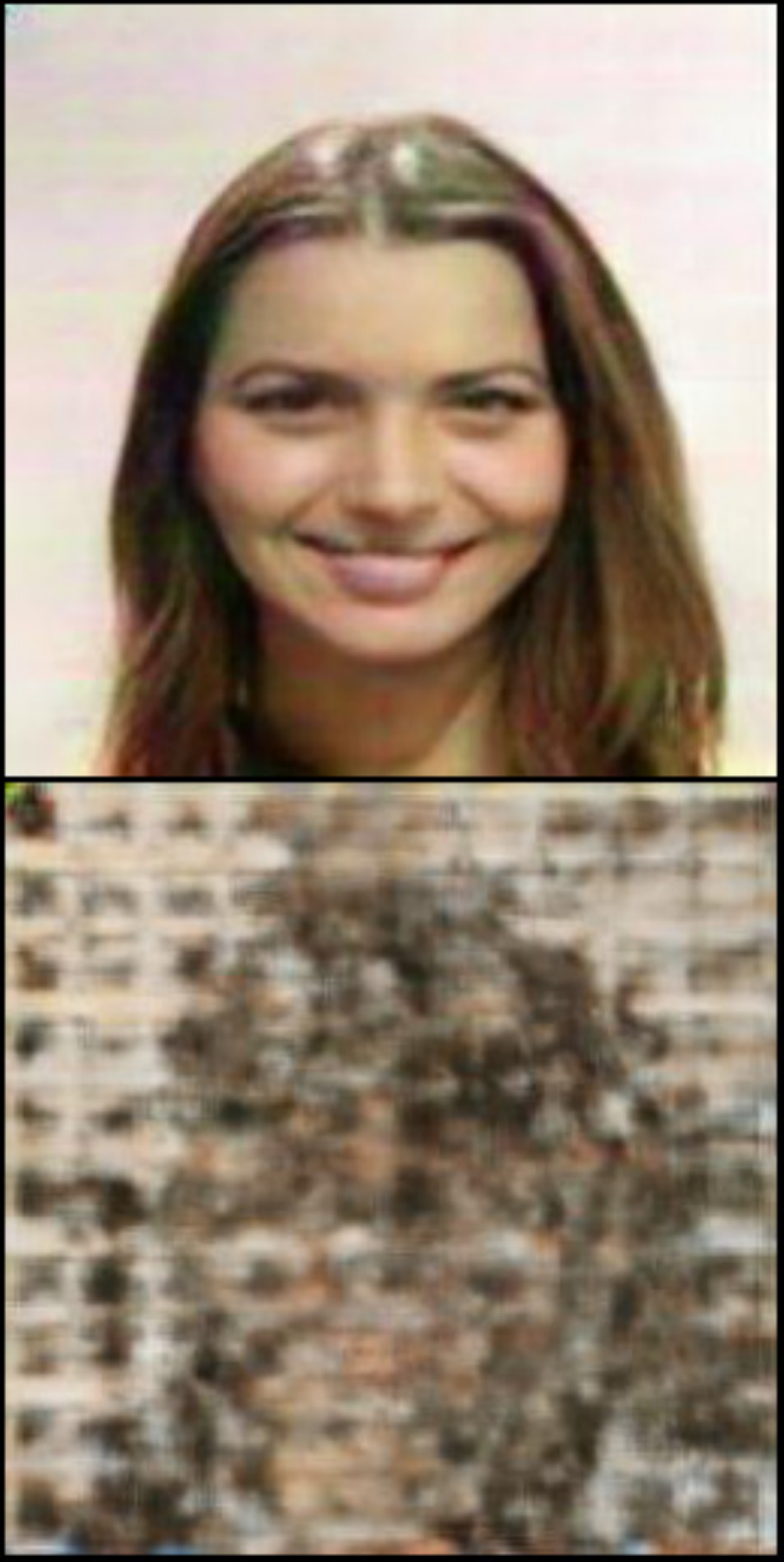}
    &   \includegraphics[width=0.12\linewidth]{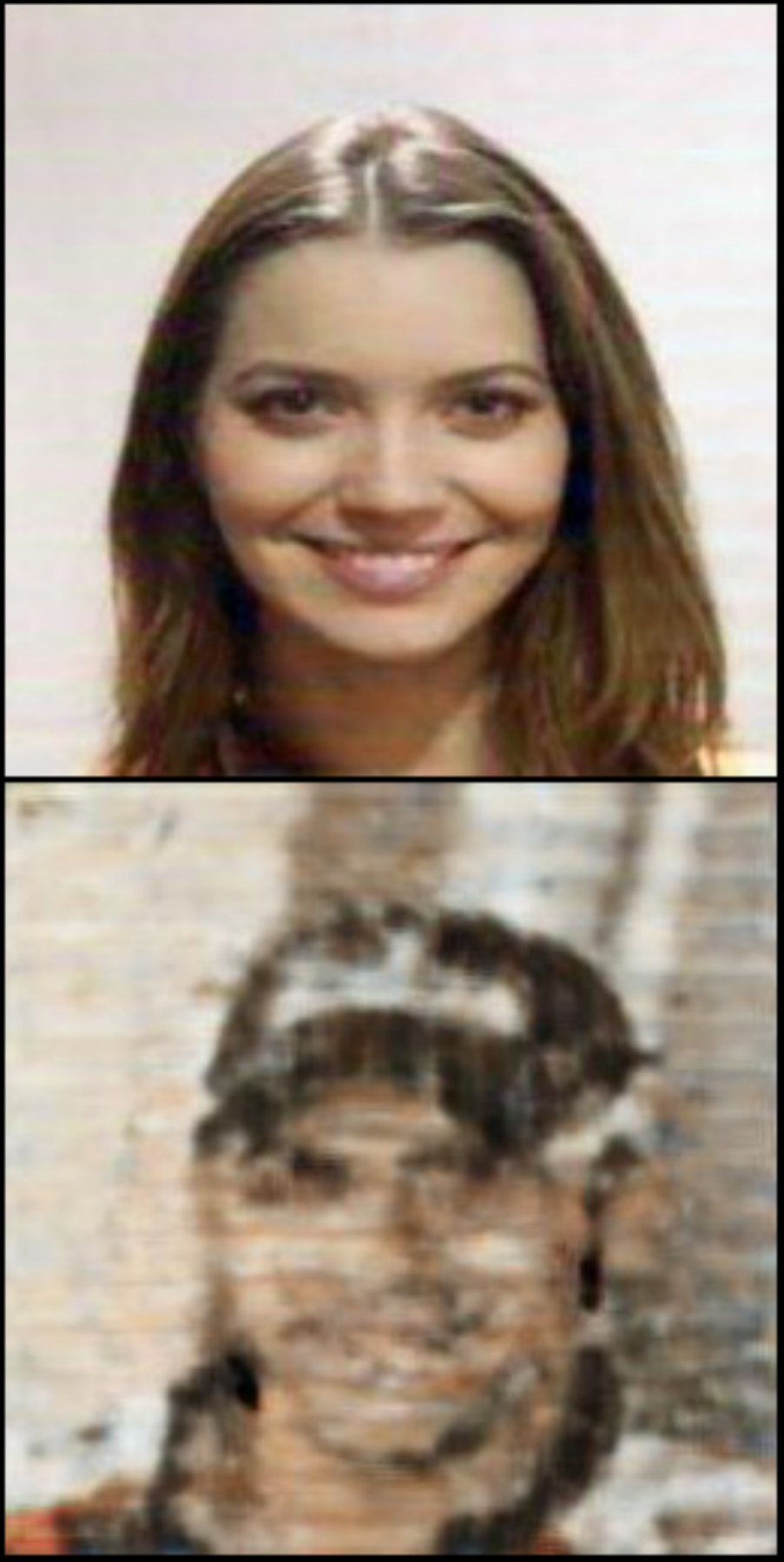}
    &   \includegraphics[width=0.12\linewidth]{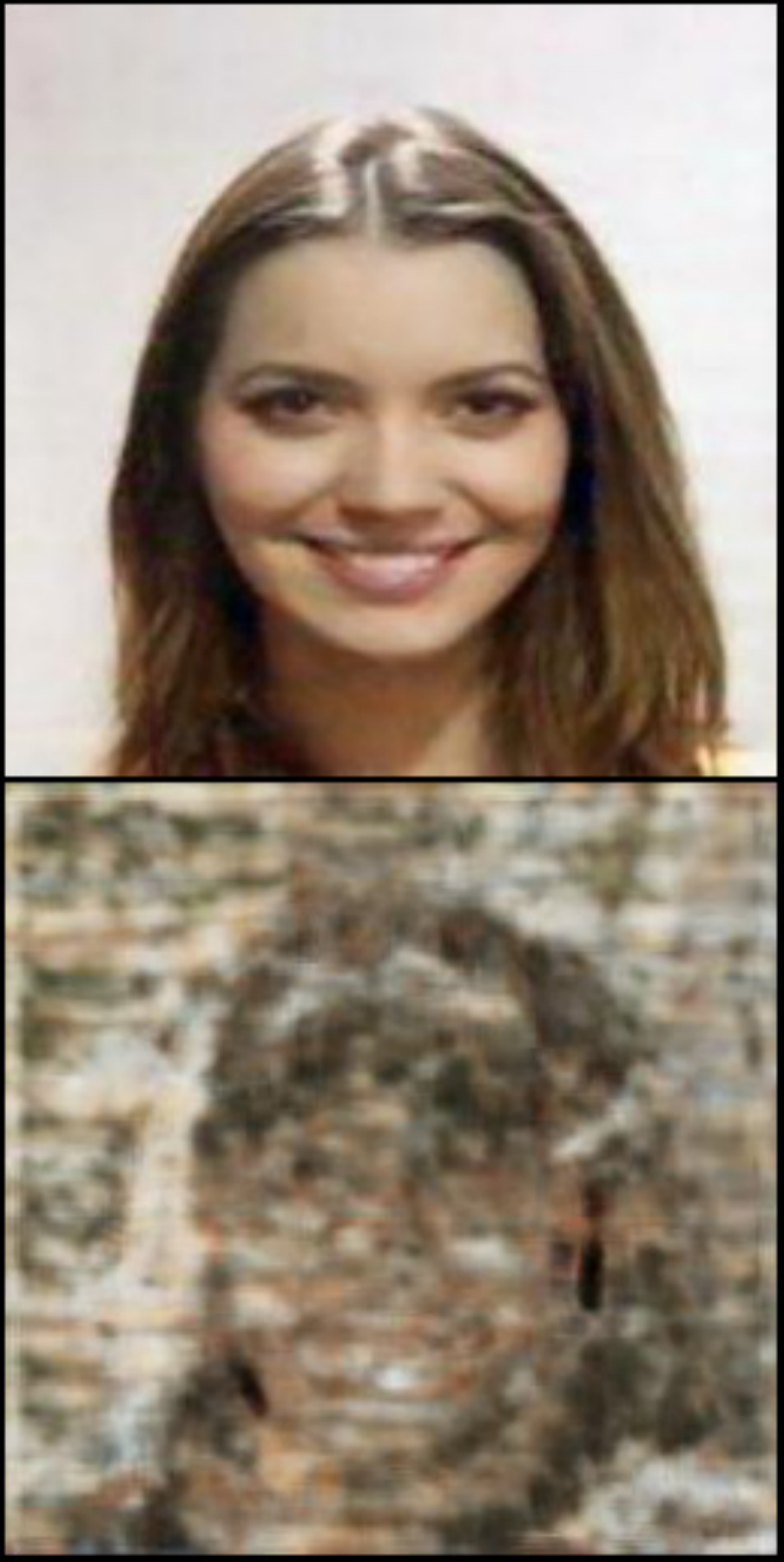}
    &   \includegraphics[width=0.12\linewidth]{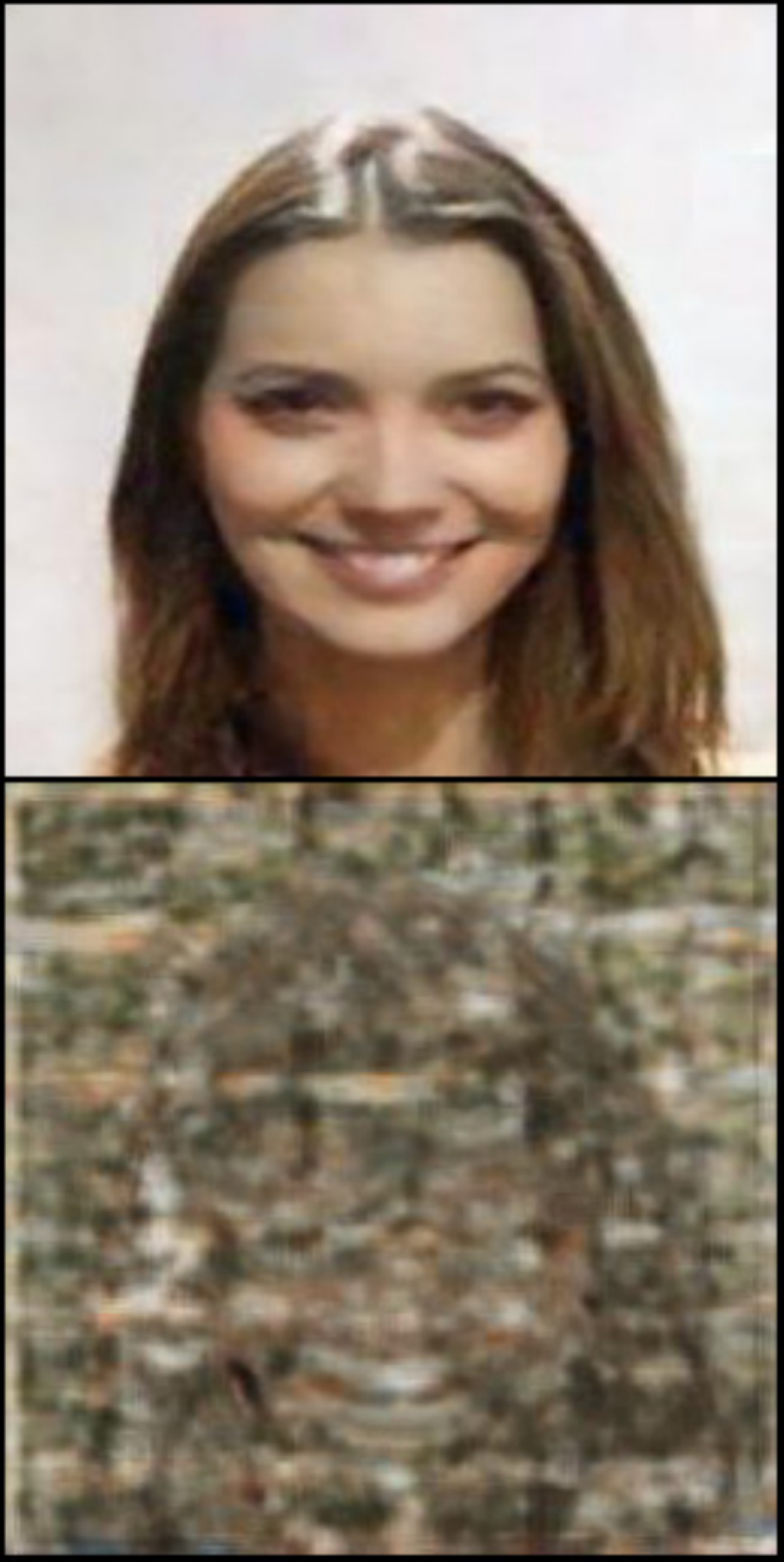}
     \\
   0.371|0.036& 0.585|0.179 &0.513|0.091& 0.444|0.036\\
    \end{tabular}%
  }
    \caption{Examples of attacking UDH-P using PEEL and PEEL-O with different $\delta$.}\label{fig:examples_udhp}%
\end{figure}

We further evaluate PEELs on UDH when the defender knows the attacker's configurations. Specifically, we train two adversarially enhanced UDH schemes. For the first one, we train a combined UDH (UDH-C) based on multiple distortions (e.g. Gaussian blur, Gaussian noise, Crop, Cropout, etc).
For the second one, we train an UDH scheme UDH-P adversarially based on PEEL-O ($\delta=0.05$). After the training processes, we use PEEL and PEEL-O with different $\delta$ to attack UDH-C and UDH-P. Table \ref{tab:udh_adv} illustrates the attack results. We observe that the VIF-S values of  PEEL and PEEL-O(0.05) on UDH-C are 0.014 and 0.015, which means no visual information is recovered. Thus, UDH-C cannot defend PEEL and PEEL-O(0.05). The VIF-S value of PEEL on UDH-P is only 0.038, which also indicates that UDH-P fails in defending PEEL. We observe that PEEL-O(0.05) cannot remove the secret embedded by UDH-P. However, as we increase the value of $\delta$ in PEEL-O, the VIF-S value decreases. When we set $\delta=0.2$, the VIF-S value is only 0.036. Although the VIF-C value also decreases to 0.402, it is still larger than corresponding VIF-S value of PEEL. We show the visualization examples in Figure \ref{fig:examples_udhp}, from which we can also conclude that our PEELs still work even the defender knowns more information about the attacking process.

\section{Conclusions}\label{sec:conclusion}
In this paper, we first explore the vulnerabilities of current deep hiding schemes. Then, according to our observations, we propose a novel provable removal attack on deep hiding to erase embedded secret images from containers. We also design two optimization strategies to improve the attack efficiency of PEEL and the quality of containers. Both theoretical and experimental results indicate that our PEELs can remove secret images thoroughly without any prior knowledge of depp hiding schemes, even the schemes are enhanced by adversarial training. We hope our work can heat up the arms race to inspire the designs of more advanced deep hiding schemes and attacks in the future.

From the defense aspect, one potential defense is to design deep hiding schemes that overcome the locality and low redundancy vulnerabilities, such as hiding each secret pixel in a more big region and improve the redundancy of each pixel. However, this may affect the message capacity.  We will consider this as an important direction for future work.

\clearpage
\bibliography{main}

\end{document}